\documentclass[11pt,reqno]{JHEP3}
\usepackage[dvips]{graphicx}
\usepackage{epsfig}
\newcommand{\mmch}[4][z,q]{\chi_{\frac{#3}{2} \frac{#4}{2}}^{k=#2} 
(#1)}
\newcommand{\Rmmch}[4][z,q]{\chi_{#3\, #4}^{k=#2} (#1)}
\newcommand{\tmmch}[4][\tilde{q}]{\tilde{\chi}_{\frac{#3}{2} 
\frac{#4}{2}}^{k=#2}(#1)}
\newcommand{\dgch}[3][z,q]{\chi_{(#2,#3)}^{c=3,\text{NS}} (#1)}
\newcommand{\Rdgch}[2]{\chi_{(#1,#2)}^{c=3,\text{R}} (z,q)}
\newcommand{\rrangle}{\rangle\!\rangle}
\newcommand{\llangle}{\langle\!\langle}
\newcommand{\tr}{\text{tr}}
\usepackage{amsmath,amssymb}
\allowdisplaybreaks[1]
\def\H{{\cal H}}
\title{Generalised N=2 permutation branes}  
\author{Stefan Fredenhagen and Matthias R.\ Gaberdiel\\ 
Institut f{\"u}r Theoretische Physik, ETH Z{\"u}rich\\
CH--8093 Z{\"u}rich, Switzerland\\ E-mail:
\email{stefan@itp.phys.ethz.ch}, \email{gaberdiel@itp.phys.ethz.ch}}
\abstract{Generalised permutation branes in products of $N=2$ minimal
models play an important role in accounting for all RR charges of
Gepner models. In this paper an explicit conformal field theory
construction of these generalised permutation branes for one simple
class of examples is given. We also comment on how this may be
generalised to the other cases.}   
\preprint{}
\begin{document}

\section{Introduction}

D-branes on Calabi-Yau manifolds play a central role for many
phenomenologically interesting string compactifications, and it is
important to understand them in detail. In the
large volume regime D-branes can be successfully described
geometrically, but this approach generically breaks down when the size
of the compactification manifold is of the order of the string
length. In this regime a microscopic formulation in terms of conformal
field theory is required.

At specific points in the moduli space of small Calabi-Yau manifolds
an explicit conformal field theory description is known. The relevant 
constructions include, in particular, toroidal orbifold constructions,
and Gepner models~\cite{Gepner:1987qi}. Gepner models are orbifolds of
tensor products of $N=2$ minimal models. They can be obtained as the
IR limit of Landau-Ginzburg models~\cite{Vafa1,Mar1,VW,W}. Via the
gauged linear sigma-model, Landau-Ginzburg models are in turn directly
related to Calabi-Yau manifolds~\cite{Witten:1993yc}. 

A certain class of supersymmetric D-branes for Gepner models can be
relatively easily constructed: these are the so-called tensor product
or Recknagel-Schomerus (RS) D-branes~\cite{Recknagel:1998sb} that
preserve the different $N=2$ superconformal algebras separately. Their
geometrical interpretation was understood in~\cite{Brunner:1999jq}. A
slight generalisation of the RS construction involves D-branes that
preserve the different $N=2$ superconformal algebras up to a
permutation, the so-called permutation
branes~\cite{Recknagel:2002qq}. Their geometrical interpretation has
been clarified in~\cite{Brunner:2005fv}. In general, however, these
two classes of constructions do not account for all the RR charges;
for example, for the A-type Gepner models, there are $31$ models (of
$147$) for which this is not the case~\cite{Caviezel:2005th}. It is
therefore an important open problem to find the conformal field theory
constructions that account for these missing charges. In this paper we
shall give a partial answer to this problem by constructing the
missing branes for one class of examples (that deals with $3$ of the
$31$ cases).
\smallskip

The new development that has sparked progress in this area is 
the recent characterisation of B-type D-branes for Landau-Ginzburg
models in terms of matrix factorisations of the superpotential. This
proposal, which was first made by Kontsevich in unpublished work, has
been supported (and physically motivated) in a number of papers 
\cite{Kapustin:2002bi,Brunner:2003dc,KL2,Lazaroiu:2003zi,
Herbst:2004ax}. It has subsequently been applied to the analysis of
D-branes on Calabi-Yau manifolds in 
\cite{Ashok:2004zb,Ashok:2004xq,Hori:2004ja,Brunner:2005fv,Enger:2005jk}. 
In particular, the factorisations corresponding to RS and permutation
branes could be identified.

Using these techniques it was shown in~\cite{Caviezel:2005th} that the
remaining RR charges in Gepner models can all be accounted for in
terms of factorisations that are generalisations of the ordinary
permutation factorisations. Furthermore, together with the usual RS
and permutation constructions, they account for a basis of the
complete quantised charge lattice. These new factorisations are based on
writing the superpotential for two minimal models of levels $k_{i}$ as
a product
\begin{equation}\label{genpermfact}
W=x^{r_{1}d}+y^{r_{2}d}=\prod_{\xi} (x^{r_{1}}-\xi y^{r_{2}}) \ ,
\end{equation} 
where $d=\gcd \{ k_{i}+2\}$, $k_{i}+2=r_{i}d$, and $\xi$ runs
through the $d^{\text{th}}$ roots of $-1$. If $r_1=r_2(=1)$, we have
the ordinary permutation case, and for $d=2$ the corresponding branes
are the resolved tensor product branes that occur when both levels are
even~\cite{Caviezel:2005th}. In all other cases, however, the
corresponding branes must describe new constructions that go
beyond the known class of maximally symmetric branes. These
`generalised permutation branes' thus arise when the shifted levels of
the two factor theories are different, but have a non-trivial common
factor $d\geq 3$.  

In this paper we shall give an explicit conformal field theory
construction of the generalised permutation branes for the simplest
example, namely for the product of two $N=2$ minimal models with
levels $k=1$ and $k=4$.\footnote{The shifted levels $1+2=3$ and
$4+2=6$ contain then $3$ as a common factor.} This case is
particularly simple since the total central charge of the theory is
$c=3$. For $c=3$ the representation theory of the $N=2$ algebra has
been studied in detail~\cite{Klemm:2003vn}, and (for free theories)
the most general $N=2$ superconformal branes are
known~\cite{Gaberdiel:2004nv}. We can thus analyse the above product
theory in terms of the diagonal $N=2$ algebra at $c=3$, and determine
the complete set of $N=2$ Ishibashi states. Using various symmetry
constraints, in particular the invariance of the boundary states under
spectral flow, we can then construct them explicitly. The resulting
branes reproduce precisely the topological open string spectrum that
is predicted by the matrix factorisation approach.  

The above product theory at $c=3$ is in fact equivalent to a
$\mathbb{Z}_6$-orbifold of a torus theory~\cite{Lerche:1989cs}. The
generalised permutation branes ($P$-branes) we constructed above
should therefore also have a description in terms of the torus
orbifold, and this is indeed the case. Among other things, this
identification confirms that the boundary states for the generalised
permutation branes we have constructed in terms of the diagonal $N=2$
theory are consistent. It also allows us to check at least
some of the factorisation constraints for them. Finally, it follows
from this analysis that the generalised permutation branes for the
$(k=1)\times (k=4)$ theory preserve in fact a large W-algebra, not
just the diagonal $N=2$ algebra. However, as we shall explain, this
property seems to be particular to the $(k=1)\times (k=4)$ case, and
does not appear to generalise. In fact, our arguments suggest that
for generic levels the tensor product of two $N=2$ algebras does not
contain any non-trivial W-algebra extension of the diagonal N=2
subalgebra. 

There is a second natural class of branes for the 
$(k=1)\times (k=4)$ theory that could have deserved the name  
`generalised  permutation branes'. These $\hat{P}$-branes also
preserve a large (but different) W-symmetry. Their spectrum is 
however different from the $P$-branes, and they correspond to a
different class of matrix factorisations that we identify. We
also show that the $\hat{P}$-branes have a very simple
description in terms of the permutation orbifold of  
the $(k=1)^{\times 3}$ theory (that is equivalent to the 
$(k=1)\times (k=4)$ theory). 
\medskip

The paper is organised as follows. In section~2 we give the explicit
construction of the $P$-branes for the example of $(k=1)\times (k=4)$,
and show that their topological spectrum agrees with 
the predictions from the matrix factorisation point of view. In
section~3, the description of the $P$-branes in terms of the torus
orbifold is given, and the factorisation constraints are analysed. We
also explain the extended W-symmetry that is preserved by these
branes, and why this property seems to be specific to the example of 
$(k=1)\times (k=4)$. Section~4 is devoted to the analysis of the
$\hat{P}$-branes, and section~5 contains our conclusions. We have
included a number of appendices in which some of the more technical
details of our constructions are explained.

\section{The $\boldsymbol{A_{1} \times A_{4}}$ model}

The simplest example where generalised permutation branes appear is
the tensor product of two A-type $N=2$ minimal models with levels
$k=1$ and $k=4$, respectively. In the Landau-Ginzburg
description this theory is given by the superpotential 
\begin{equation}
W = x^3 + y^6 \ ,
\end{equation}
whereas in conformal field theory it corresponds to the spectrum 
\begin{eqnarray}\label{k4k1spec}
&& \bigoplus_{[l_i,m_i,s_i]} 
\Bigl(\H_{[l_1,m_1,s_1]} \otimes \H_{[l_2,m_2,s_2]}  \Bigr)
\otimes
\Bigl( \bar{\H}_{[l_1,m_1,-s_1]} \otimes \bar{\H}_{[l_2,m_2,-s_2]} 
\nonumber \\ 
&& \qquad \qquad \qquad \qquad \qquad \qquad \qquad \qquad 
\oplus \, 
\bar{\H}_{[l_1,m_1,-s_1+2]} \otimes \bar{\H}_{[l_2,m_2,-s_2+2 ]}
\Bigr)  \ .
\end{eqnarray}
Here the sum runs only over those pairs of representations for
which $s_1-s_2$ is even, and $[l_1,m_1,s_1]$ ($[l_2,m_2,s_2]$)
denotes the representations of the bosonic subalgebra of the $N=2$
superconformal algebra at $k=1$ ($k=4$); for a description of the
usual conventions see appendix~A  (compare also 
\cite{Recknagel:1998sb,Fuchs:2000gv,Brunner:2005fv}).

\subsection{Branes in matrix factorisations}

Let us first briefly review what information about the generalised
permutation branes can be obtained from the matrix factorisation
approach. We begin by collecting some basic facts about the relation
between matrix factorisations and D-branes. 

According to Kontsevich's proposal, B-type D-branes in
Landau-Ginzburg models correspond to matrix factorisations of the
superpotential $W$,  
\begin{equation}\label{factorize}
E\, J = J \, E = W\cdot {\bf 1}\ ,
\end{equation}
where $E$ and $J$ are $r\times r$ matrices with polynomial entries in
the superfields. This condition can be more succinctly written as  
\begin{equation} \label{Qdef}
Q^2=W\cdot \mathbf{1}\ , \qquad \mbox{where} \qquad 
Q=\left(\begin{array}{cc} 0 & J \\ E & 0 \end{array}\right) \ . 
\end{equation}
The matrices $E$ and $J$ describe boundary fermions whose presence is 
required to cancel the supersymmetry variation of the bulk
F-term in the presence of a boundary. This approach was proposed in
unpublished form by Kontsevich, and the  physical interpretation of it
was given in~\cite{Kapustin:2002bi,Brunner:2003dc,KL2};  for a good
review of this material see for example~\cite{Hori:2004zd}. 

Given the close relation between Landau-Ginzburg models and $N=2$
superconformal field theories, we therefore expect that there is a
one-to-one correspondence between matrix factorisations and $N=2$
superconformal D-branes. Furthermore, the topological part of the open
string spectrum between any two such D-branes should correspond to a
suitable cohomology that can be directly calculated in terms of the
matrix factorisation description. This will allow us to match matrix
factorisations with superconformal D-branes in conformal field
theory. 
\smallskip

For the case of the above Landau-Ginzburg model we have the obvious tensor
product factorisations   
\begin{equation}
E = \left(\begin{matrix} E_1& E_2\\ J_2 & -J_1\end{matrix}\right) \ ,
\qquad
J = \left( \begin{matrix} J_1 & E_2 \\ J_2 & - E_1
\end{matrix}\right)  \ ,
\end{equation}
where $E_1=x^l$, $J_1=x^{3-l}$,  $E_2=y^m$, $J_2=y^{6-m}$. Up to the
usual equivalences we may restrict ourselves to $l=1$ and $m=1,2,3$.  

These factorisations however do not carry any RR charge. The simplest
factorisation that carries RR charge is the generalised permutation
factorisation that was first considered 
in~\cite{Caviezel:2005th}. 
In order to see how it arises one observes that $W$ can be written as 
\begin{equation}\label{genpermfactexample}
W = (x -\xi_0 y^{2})\, ( x - \xi_1 y^{2})\, (x - \xi_2 y^{2}) \ , 
\qquad
\xi_j = e^{\frac{(2j+1) i\pi}{3}} \ .
\end{equation}
A rank $1$ factorisation of $W=EJ$ is then obtained by taking $E$ to
be one of the three factors above, while $J$ is the  
product of the other two. There are three such choices
that we shall denote by $Q_\xi$ (where $\xi$ denotes the root that
appears in the factor $E$), and together with their reverse
factorisations $Q^r_\xi$ (where the roles of $E$ and 
$J$ are interchanged) we have in total six such factorisations. 

It is the aim of this paper to identify the corresponding branes (that
we shall call '$P$-branes' in the following) in conformal field theory.
Before we start to construct the superconformal boundary states we
shall collect some information on the spectrum of these D-branes from
the matrix factorisation point of view. This information will help us
identify these branes correctly in conformal field theory.

\subsubsection{Spectra from matrix factorisations}

It is straightforward to calculate the topological open string
spectrum between any two such 
factorisations. By definition, the spectrum between two factorisations
$Q_1$ and $Q_2^r$ can be obtained from that between $Q_1$ and $Q_2$ by
exchanging the roles of the bosons and fermions. It is therefore
sufficient to give the results only for the spectrum between the
factorisations $Q_\xi$. For the case at hand, the results of 
\cite{Caviezel:2005th} can be summarised as follows:
\smallskip

{\bf (1)}: The self-spectrum of $Q_\xi$ with itself contains 
no fermions and four bosons of $U(1)$ charge
$0,\tfrac{1}{3},\tfrac{2}{3},1$. 
\smallskip

{\bf (2)}: The spectrum between $Q_\xi$ and $Q_{\xi'}$, where 
$\xi\ne \xi'$, contains no bosons, and two fermions of $U(1)$ charge 
$\tfrac{1}{3},\tfrac{2}{3}$. 
\smallskip

{\bf (3)}: The relative spectrum between the tensor product
factorisations corresponding to $E_1=x$, $E_2=y$ and $Q_\xi$ is
independent of $\xi$. It contains one boson of $U(1)$ charge
$\tfrac{2}{3}$ and one fermion of $U(1)$ charge $\tfrac{1}{3}$. 
\smallskip

\noindent After these preparations we shall now construct the
corresponding boundary states in conformal field theory.

\subsection{Branes in conformal field theory}

The branes that correspond to matrix factorisations of the
Landau-Ginzburg model are superconformal B-type D-branes of the
diagonal $N=2$ algebra. The boundary states $|\!|B\rrangle$ of B-type
D-branes are characterised by the condition that
\begin{eqnarray}
\left(L_n - \tilde{L}_{-n} \right) |\!| B \rangle\!\rangle & = & 0
\nonumber \\
\left(J_n + \tilde{J}_{-n} \right) |\!| B \rangle\!\rangle & = & 0
\label{gluing} \\
\left(G^\pm_r + i\, \eta\, \tilde{G}^\pm_{-r} \right) 
|\!| B \rangle\!\rangle & = & 0 \ , \nonumber
\end{eqnarray}
where $\eta=\pm$ distinguishes the two different spin-structures. In
the following we shall work with one fixed choice of $\eta$, say
$\eta=+1$.  

The theory in question possesses however more symmetry since we have a 
tensor product of two $N=2$ algebras with $c=1$ ($k=1$) and $c=2$
($k=4$), respectively. One simple class of branes are those that
respect the full chiral symmetry
\begin{eqnarray}
\left(L^{(1)}_n - \tilde{L}^{(1)}_{-n} \right) |\!| B \rangle\!\rangle
=  \left(L^{(2)}_n - \tilde{L}^{(2)}_{-n} \right) |\!| B \rangle\!\rangle
& = & 0 \nonumber \\
\left(J^{(1)}_n + \tilde{J}^{(1)}_{-n} \right) |\!| B \rangle\!\rangle 
= \left(J^{(2)}_n + \tilde{J}^{(2)}_{-n} \right) |\!| B \rangle\!\rangle 
& = & 0 \label{gluingten} \\
\left(G^{(1)\pm}_r + i\,  \tilde{G}^{(1)\pm}_{-r} \right) 
|\!| B \rangle\!\rangle =
\left(G^{(2)\pm}_r + i\,  \tilde{G}^{(2)\pm}_{-r} \right) 
|\!| B \rangle\!\rangle 
& = & 0 \ . \nonumber
\end{eqnarray}
These `tensor product branes' (or 
Recknagel-Schomerus branes \cite{Recknagel:1998sb}) are labelled 
by a  qua\-druple $(L_{1},L_{2},S_{1},S_{2})$ of integers where
$L_{1}=0,1$ and $L_{2}=0,\dotsc ,4$, and $S_{1},S_{2}$ are defined
modulo $4$. The  difference $S_{1}-S_{2}$ is always even and there are
the identifications
\begin{equation}
(L_{1},L_{2},S_{1},S_{2}) \sim (1-L_{1},L_{2},S_{1}+2,S_{2}) \sim
(L_{1},4-L_{2},S_{1},S_{2}+2) \sim (L_{1},L_{2},S_{1}+2,S_{2}+2)
\end{equation}
among the quadruples, so we can always choose $L_{1}=0$ and
$L_{2}=0,1,2$. The choice $\eta=+1$ in (\ref{gluing}) corresponds now
to considering only branes with even $S_{1}$ and $S_{2}$. Given the
above identifications, there are only two cases, namely $S_1=S_2=0$
and $S_1=0, S_2=2$, which are anti-branes of one another. It is
therefore sufficient to give only the branes 
$|\!|T(L_{1},L_{2})\rrangle $ with $S_1=S_2=0$, for which we have
\begin{align}
|\!|T (0,0)\rrangle \ & = \ \frac{3^{1/4}}{\sqrt{2}}\big(
|\tfrac{1}{2}\tfrac{1}{2},\tfrac{1}{2}\tfrac{1}{2}\rrangle
+\sqrt{2}|\tfrac{1}{2}\tfrac{1}{2},\tfrac{3}{2}\tfrac{3}{2}\rrangle +
|\tfrac{1}{2}\tfrac{1}{2},\tfrac{5}{2} \tfrac{5}{2}\rrangle
\big)+\sqrt{\frac{3}{2}} \big(|1\,1,1\,1\rrangle + |1\,1,2\,2\rrangle
\big) \\ 
|\!|T (0,1)\rrangle \ &=\ \frac{3^{\frac{3}{4}}}{\sqrt{2}} \big(
|\tfrac{1}{2}\tfrac{1}{2},\tfrac{1}{2}\tfrac{1}{2}\rrangle -
|\tfrac{1}{2}\tfrac{1}{2}, \tfrac{5}{2}\tfrac{5}{2}\rrangle\big)
+\sqrt{\frac{3}{2}} \big(|1\,1,1\,1\rrangle - |1\,1,2\,2\rrangle\big) \\
|\!|T (0,2)\rrangle \ &=\ 3^{\frac{1}{4}} \big(
\sqrt{2}|\tfrac{1}{2}\tfrac{1}{2},\tfrac{1}{2}\tfrac{1}{2}\rrangle -
|\tfrac{1}{2}\tfrac{1}{2},\tfrac{3}{2}\tfrac{3}{2} \rrangle
+\sqrt{2}|\tfrac{1}{2}\tfrac{1}{2},\tfrac{5}{2}\tfrac{5}{2}\rrangle \big) \ .
\end{align}
Note that $T (0,2)$ is its own anti-brane and does
not couple to the RR sector. The states that appear on the right hand
side are the B-type tensor product Ishibashi states. Our conventions
for the labelling of the representations of the $N=2$ superconformal
algebra are explained in appendix~A. 

These tensor product branes correspond to the tensor product
factorisations that were described at the beginning of section~2.1 
\cite{Brunner:2003dc,KL2,Kapustin:2003rc}. In fact, the factorisations
with  $E_1=x$, $E_2=y^m$ correspond to $|\!|T(0,m-1)\rrangle$.  

\subsubsection{Generalised permutation branes and spectral flow} 

The tensor product branes are the only branes that preserve the full
chiral symmetry. In particular, since the central charges of the two
$N=2$ minimal models are different, it is not possible to construct
the usual `permutation branes' \cite{Recknagel:2002qq} (see also
\cite{Gaberdiel:2002jr,Sarkissian:2003yw}) that are
characterised by the gluing conditions
\begin{eqnarray}
\left(L^{(1)}_n - \tilde{L}^{(2)}_{-n} \right) |\!| B \rangle\!\rangle 
= \left(L^{(2)}_n - \tilde{L}^{(1)}_{-n} \right) |\!| B \rangle\!\rangle
& = & 0 \nonumber \\
\left(J^{(1)}_n + \tilde{J}^{(2)}_{-n} \right) |\!| B \rangle\!\rangle 
= \left(J^{(2)}_n + \tilde{J}^{(1)}_{-n} \right) |\!| B \rangle\!\rangle 
& = & 0 \label{gluingperm} \\
\left(G^{(1)\pm}_r + i\,  \tilde{G}^{(2)\pm}_{-r} \right) 
|\!| B \rangle\!\rangle =
\left(G^{(2)\pm}_r + i\,  \tilde{G}^{(1)\pm}_{-r} \right) 
|\!| B \rangle\!\rangle & = & 0 \ . \nonumber
\end{eqnarray}
On the other hand, we have seen that there exists an additional simple
class of factorisations, namely the generalised permutation 
factorisations~\eqref{genpermfactexample} that are very reminiscent
of the factorisations that correspond to the usual permutation
branes~\cite{Brunner:2005fv}. We should therefore expect that the
corresponding boundary states are a simple generalisation of the
permutation brane construction.

In order to make progress with this construction we use the fact that
the relevant branes must respect the diagonal $N=2$ symmetry. The two
individual $N=2$ algebras have $c=1$ and $c=2$, respectively, and thus
the diagonal $N=2$ algebra has $c=3$. Its representation theory was
analysed in detail in \cite{Klemm:2003vn,Gaberdiel:2004nv}. It is not
difficult to decompose the various tensor products of $N=2$
representations in terms of the diagonal $N=2$ algebra; the relevant
decompositions are given in appendix~B.

Since we are considering the diagonal modular invariant
(\ref{k4k1spec}), we have one B-type Ishibashi state of the
diagonal $N=2$ algebra for each $N=2$ representation that appears
together with its conjugate representation in the decomposition of the
tensor product of the two minimal model representations.\footnote{For
the RR sector one also has to keep track of the constraints imposed by
the GSO-projection.} In particular, it is therefore clear from the
results of appendix~B that there are infinitely many
B-type Ishibashi states for the diagonal $N=2$ algebra, and thus a
continuum of B-type D-branes. We expect that the D-branes that
correspond to the generalised permutation factorisations are rather
special. In order to identify them we need to use additional
constraints. One useful constraint comes from the analysis of the
spectral flow symmetry.

It is easy to see that the B-type gluing conditions are compatible
with applying the spectral flow
automorphism~\eqref{spectralflowofalgebra} for the left- and
right-movers with opposite values of $t$. The spectral flow
symmetry of boundary states is intimately connected to the
quantisation of the $U (1)$-charge in the open string
spectrum. Indeed, if the B-type boundary states $|\!|B\rrangle$ and
$|\!|B'\rrangle$ are invariant under a spectral flow by $t$
($-t$) on the left-(right-)movers, then their overlap satisfies
\begin{equation}
\llangle B'|\!|q^{\frac{1}{2} (L_{0}+\tilde{L}_{0})-\frac{c}{24}}
z^{\frac{1}{2} (J_{0}-\tilde{J}_{0})} |\!| B\rrangle  =
\llangle B'|\!|q^{\frac{1}{2}
(L_{0}+\tilde{L}_{0})-\frac{c}{24}+t^{2}\frac{c}{6}
-t \frac{1}{2} (J_{0}-\tilde{J}_{0})} 
z^{\frac{1}{2} (J_{0}-\tilde{J}_{0})-t \frac{c}{3}}|\!|B\rrangle \ .
\end{equation}
By modular transformation, this translates into a relation for the
open string partition function,
\begin{equation}
\tr_{\mathcal{H}_{BB'}}\big(\tilde{q}^{L_{0}-\frac{c}{24}}
\tilde{z}^{J_{0}} \big)  = \tr_{\mathcal{H}_{BB'}}\big(
\tilde{q}^{L_{0}-\frac{c}{24}} \tilde{z}^{J_{0}}e^{-2\pi it J_{0}} 
\big)  \ .
\end{equation}
This then implies that the $U (1)$-charges $q$ of open strings between
$|\!|B\rrangle$ and $|\!|B'\rrangle$ are quantised and satisfy 
$t q\in \mathbb{Z}$. 

From the Landau-Ginzburg description we know that the topological part
of the open string spectrum between the generalised permutation 
factorisations only contains $U (1)$-charges which are multiples of
$\frac{1}{3}$. This suggests that the corresponding boundary states
are in fact invariant under spectral flow by $t =3$. This is
further supported by the observation that the tensor product branes
$|\!|T (0,0)\rrangle$ and $|\!|T(0,2)\rrangle$ are invariant
under this spectral flow, and that the topological part of the
$U(1)$-charges in the relative spectrum between these and the 
generalised permutation factorisations are also quantised in units of 
$\frac{1}{3}$. The boundary state $|\!|T (0,1)\rrangle$ on the other
hand, picks up a minus sign under the spectral flow by $t =3$, and
the relative spectrum between this brane and the generalised
permutation branes should therefore only have 
$U(1)$-charges satisfying $q\in\frac{1}{6}+\frac{1}{3}\mathbb{Z}$. At
least for the topological part of the spectrum this is indeed the
case. 

The requirement that the boundary states are invariant under spectral
flow by $t =3$ restricts their structure significantly. For
example, under $t=\pm 3$ the representations of the $k=1$ factor are
invariant, but for the $k=4$ theory we exchange
\begin{equation}
t=\pm 3:   \qquad 
\left(\frac{1}{2},\frac{1}{2} \right) \leftrightarrow
\left(\frac{5}{2},\frac{5}{2} \right)  \qquad
\left(\frac{3}{2},\frac{7}{2} \right) \leftrightarrow
\left(\frac{9}{2},\frac{1}{2} \right) \qquad
\left(\frac{7}{2},\frac{3}{2} \right) \leftrightarrow
\left(\frac{1}{2},\frac{9}{2} \right) \ ,
\end{equation}
while the representations $(\tfrac{3}{2},\tfrac{3}{2})$,
$(\tfrac{5}{2},\tfrac{1}{2})$ and $(\tfrac{1}{2},\tfrac{5}{2})$ are
invariant (see appendix~A). Thus the spectral flow symmetry fixes for
example the coefficients of the B-type Ishibashi states that appear in
the sector $(\tfrac{1}{2}\tfrac{1}{2},\tfrac{5}{2}\tfrac{5}{2})$ in
terms of those appearing in
$(\tfrac{1}{2}\tfrac{1}{2},\tfrac{1}{2}\tfrac{1}{2})$, {\it
etc}. Furthermore, since the spectral flow of the $c=3$ theory by
$t=\pm 3$ relates the representations of the form
$(6n+\tfrac{5}{2},1) \leftrightarrow (6n+\tfrac{11}{2},1)$, we also
know that the coefficients with which the corresponding Ishibashi
states appear must be the same.

There is also a spectral flow symmetry in the open string spectrum
which is the result of a charge quantisation in the boundary
states. In the theory we are considering, B-type branes can only
couple to states with integer $U(1)$ charge, so that all open string
spectra are invariant under a spectral flow by $t =\pm 1$. As this
symmetry is a consequence of the B-type condition it does not give 
additional information on the boundary states, but being aware of this
symmetry helps to understand the open string spectra by organising
them into spectral flow orbits.

While these considerations constrain the form of the boundary state,
they do not determine it uniquely. The construction of the boundary
state therefore requires a certain amount of guess work. In the
following we shall make an ansatz for the boundary state; we
shall then give various pieces of evidence that suggest that this is
indeed the correct choice. In particular, we shall show that
it reproduces the correct topological open string spectra. In section
3 we shall also show that it has a natural interpretation in terms of
a free field construction.

\subsubsection[The construction of the $P$-branes in the 
NSNS sector]{The construction of the $\boldsymbol{P}$-branes in the
NSNS sector}  

We shall first restrict our discussion to the NSNS sector. As we have
just explained, the spectral flow symmetry restricts the possible
couplings to the different $N=2$ Ishibashi states. We now make the
ansatz that the generalised permutation branes couple to the following
linear combinations of Ishibashi states:
\begin{align}
|\tfrac{1}{2}\tfrac{1}{2},\tfrac{1}{2}\tfrac{1}{2}\rrangle^{P}\
=\ & |0,0\rrangle  + \sum_{m=1}^{\infty}\sum_{n=1}^{m} \tfrac{1}{3}
\big((-1)^{m}+ (-1)^{n}+ (-1)^{m+n} \big)  
|m^{2}+n^{2}-mn,0\rrangle  \nonumber \\
& \qquad + \sum_{n=0}^{\infty} \big( |6n+\tfrac{11}{2},1\rrangle +
|6n+\tfrac{11}{2},-1\rrangle \big) \label{vacstr}\\[1mm]
|\tfrac{1}{2}\tfrac{1}{2},\tfrac{5}{2}\tfrac{5}{2}\rrangle^{P}\
=\ & \sum_{m=1}^{\infty}\sum_{n=1}^{m}\tfrac{1}{3}
\big((-1)^{m}+ (-1)^{n}+ (-1)^{m+n} \big) 
|m^{2}+n^{2}-mn,0\rrangle  \nonumber \\
& \qquad + \sum_{n=0}^{\infty} \big(
|6n+\tfrac{5}{2},1\rrangle + |6n+\tfrac{5}{2},-1\rrangle  \big)\\[1mm]
|\tfrac{1}{2}\tfrac{1}{2},\tfrac{3}{2}\tfrac{3}{2}\rrangle^{P}\
=\ & \sum_{m=0}^{\infty}\sum_{n=0}^{\infty}
\tfrac{1}{3} \big( (-1)^{m+1}+ (-1)^{n}+ (-1)^{m+n+1}\big)
|\tfrac{1}{3}+ (m-n)^{2}+ (m+1)n,0\rrangle \\[1mm]
|\tfrac{3}{2}\tfrac{1}{2},\tfrac{3}{2}\tfrac{7}{2}\rrangle^{P}\ =\ & 
\sum_{m=1}^{\infty}\sum_{\substack{n=1\\
\makebox[0cm][c]{$\scriptstyle\!\!\!\!\!\text{$m$ or $n$ odd}$}}}^{m} 
\tfrac{2}{3}
|m^{2}+n^{2}-mn,0\rrangle \\[1mm]
|\tfrac{3}{2}\tfrac{1}{2},\tfrac{9}{2}\tfrac{1}{2}\rrangle^{P}\ =\ & 
\sum_{m=1}^{\infty}\sum_{\substack{n=1\\
\makebox[0cm][c]{$\scriptstyle\!\!\!\!\!\text{$m$ or $n$ odd}$}}}^{m} 
\tfrac{2}{3}
|m^{2}+n^{2}-mn,0\rrangle \\[1mm]
|\tfrac{1}{2}\tfrac{3}{2},\tfrac{7}{2}\tfrac{3}{2}\rrangle^{P}\ =\ & 
\sum_{m=1}^{\infty}\sum_{\substack{n=1\\
\makebox[0cm][c]{$\scriptstyle\!\!\!\!\!\text{$m$ or $n$ odd}$}}}^{m} 
\tfrac{2}{3}
|m^{2}+n^{2}-mn,0\rrangle \\[1mm]
|\tfrac{1}{2}\tfrac{3}{2},\tfrac{1}{2}\tfrac{9}{2}\rrangle^{P}\ =\ & 
\sum_{m=1}^{\infty}\sum_{\substack{n=1\\
\makebox[0cm][c]{$\scriptstyle\!\!\!\!\!\text{$m$ or $n$ odd}$}}}^{m} 
\tfrac{2}{3}
|m^{2}+n^{2}-mn,0\rrangle\\[1mm]
|\tfrac{3}{2}\tfrac{1}{2},\tfrac{1}{2}\tfrac{5}{2}\rrangle^{P}\ =\ &
\sum_{m=0}^{\infty}\sum_{\substack{n=0\\
\makebox[0cm][c]{$\scriptstyle\!\!\!\!\!\text{$m+1$ or $n$
    odd}$}}}^{\infty} 
\tfrac{2}{3} | \tfrac{1}{3}+ (m-n)^{2}+ (m+1)n,0\rrangle \\[1mm]
|\tfrac{1}{2}\tfrac{3}{2},\tfrac{5}{2}\tfrac{1}{2}\rrangle^{P}\ =\ &
\sum_{m=0}^{\infty}\sum_{\substack{n=0\\
\makebox[0cm][c]{$\scriptstyle\!\!\!\!\!\text{$m+1$ or $n$
    odd}$}}}^{\infty} 
\tfrac{2}{3} | \tfrac{1}{3}+ (m-n)^{2}+ (m+1)n,0\rrangle
 \ .
\end{align}
We have chosen here the convention that the combination of $N=2$
Ishibashi states denoted by $|u_1v_1,u_2,v_2\rrangle^P$ describes the 
B-type Ishibashi states from the representation 
$(u_1,v_1)\otimes (u_2,v_2)$ of the product theory (for the
decomposition of the tensor product representations see
appendix~B.1). On the right hand side, the $N=2$ Ishibashi states are
denoted by their weight and charge (with respect to the diagonal $N=2$
algebra). We have not distinguished the different $N=2$ Ishibashi
states with the same weight and charge; for example there are six
different Ishibashi states $|m^2+n^2-mn,0\rrangle$ for each allowed
combination of $m$ and $n$ that appear in six different sectors. In
addition, different combinations of $m$ and $n$ are understood to
correspond to different Ishibashi states even if their conformal
weight and charge coincide. For the first three sectors (for which we
also have tensor product Ishibashi states) we have furthermore chosen
the convention that the tensor product Ishibashi states involve all
relevant $N=2$ Ishibashi states with (relative) coefficient $1$.

With these notations we now claim that the boundary states for the
generalised permutation branes are ($j=0,1,2$) 
\begin{align}
|\!|P (j)\rrangle_{\text{NSNS}} \ =\ & \frac{3^{\frac{1}{4}}}{\sqrt{2}}
\bigg(|\tfrac{1}{2}\tfrac{1}{2},\tfrac{1}{2}\tfrac{1}{2}\rrangle^{P}
+|\tfrac{1}{2}\tfrac{1}{2},\tfrac{5}{2}\tfrac{5}{2}\rrangle^{P}
+\sqrt{2}
|\tfrac{1}{2}\tfrac{1}{2},\tfrac{3}{2}\tfrac{3}{2}\rrangle^{P}\nonumber\\
& \qquad  +e^{\frac{2\pi ij}{3}}\Big(
|\tfrac{3}{2}\tfrac{1}{2},\tfrac{3}{2}\tfrac{7}{2}\rrangle^{P}  
+|\tfrac{3}{2}\tfrac{1}{2},\tfrac{9}{2}\tfrac{1}{2}\rrangle^{P}
+\sqrt{2}
|\tfrac{3}{2}\tfrac{1}{2},\tfrac{1}{2}\tfrac{5}{2}\rrangle^{P} 
\Big)\nonumber\\
& \qquad  +e^{-\frac{2\pi ij}{3}}\Big(
|\tfrac{1}{2}\tfrac{3}{2},\tfrac{7}{2}\tfrac{3}{2}\rrangle^{P}  
+|\tfrac{1}{2}\tfrac{3}{2},\tfrac{1}{2}\tfrac{9}{2}\rrangle^{P}
+\sqrt{2}
|\tfrac{1}{2}\tfrac{3}{2},\tfrac{5}{2}\tfrac{1}{2}\rrangle^{P} \Big)
\bigg)\ .
\end{align}
This proposal respects in addition the $\mathbb{Z}_{3}$-symmetry under
which the three $P$-branes are connected: in the Landau-Ginzburg 
description this is the symmetry $x\to e^{\frac{2\pi i}{3}}x$, under
which $Q_{\xi_{i}}$ transforms into $Q_{\xi_{i-1}}$. In conformal
field theory this corresponds to multiplying 
the contribution from a sector with the coset labels
$(l_{1},m_{1},s_{1})\otimes (l_{2},m_{2},s_{2})$ by the 
phase $e^{\frac{2\pi im_{1}}{3}}$.

\subsubsection[The construction of the $P$-branes in the RR sector]{The 
construction of the $\boldsymbol{P}$-branes in the RR sector}

The analysis in the RR sector is similar. The tensor product branes only
couple to the sectors with $m_{1}=m_{2}=0$, namely 
to the sectors $(1,1)\otimes (1,1)$ and $(1,1)\otimes (2,2)$. In
particular, the tensor product branes do not couple to any RR ground states.
The generalised permutation branes on the other hand couple to all
allowed RR sectors. Let us introduce the following notation for the
combination of Ishibashi states that are relevant for the $P$-branes: 
\begin{align}
|1\,0,4\,0\rrangle^{P}  = & 
|\tfrac{1}{8},0\rrangle + \sum_{n=0}^{\infty}
\big(|\tfrac{1}{8}+3+3n,\tfrac{1}{2}\rrangle  +
|\tfrac{1}{8}+3+3n,-\tfrac{1}{2}\rrangle  \big) \\
|2\,0,2\,0\rrangle^{P}  = & 
|\tfrac{1}{8},0\rrangle + \sum_{n=0}^{\infty}
\big(|\tfrac{1}{8}+3+3n,\tfrac{1}{2}\rrangle  +
|\tfrac{1}{8}+3+3n,-\tfrac{1}{2} \rrangle \big) \\
|2\,0,3\,1\rrangle^{P}  = & 
\sum_{n=0}^{\infty}
\big(|\tfrac{5}{8}+1+3n,\tfrac{1}{2}\rrangle  +
|\tfrac{5}{8}+1+3n,-\tfrac{1}{2}\rrangle  \big) \\
|1\,0,1\,3\rrangle^{P}  = & 
\sum_{n=0}^{\infty}
\big(|\tfrac{5}{8}+1+3n,\tfrac{1}{2}\rrangle  +
|\tfrac{5}{8}+1+3n,-\tfrac{1}{2}\rrangle  \big) \\
|1\,1,2\,2\rrangle^{P}  = &
\sum_{n=0}^{\infty} \big( - |\tfrac{1}{8}+1+3n,\tfrac{1}{2}\rrangle +
|\tfrac{1}{8}+1+3n,-\tfrac{1}{2}\rrangle \nonumber\\
& \qquad +
|\tfrac{1}{8}+2+3n,\tfrac{1}{2}\rrangle -
|\tfrac{1}{8}+2+3n,-\tfrac{1}{2}\rrangle \big) \\
|1\,1,1\,1\rrangle^{P}  = &
\sum_{n=0}^{\infty} \big( |\tfrac{5}{8}+3n,\tfrac{1}{2}\rrangle -
|\tfrac{5}{8}+3n,-\tfrac{1}{2}\rrangle \nonumber\\
& \qquad -
|\tfrac{5}{8}+2+3n,\tfrac{1}{2}\rrangle +
|\tfrac{5}{8}+2+3n,-\tfrac{1}{2}\rrangle \big) \ .
\end{align}
The RR part of the generalised permutation boundary states is then 
\begin{align}
|\!|P (j)\rrangle_{\text{RR}} \ =\ & \frac{i}{\sqrt{2}} \big(
|1\,1,1\,1\rrangle^{P}  + |1\,1,2\,2\rrangle^{P}  \big) \nonumber\\
&+ e^{\frac{2\pi ij}{3}}\big( |1\,0,4\,0\rrangle^{P}
+|1\,0,1\,3\rrangle^{P}\big)\nonumber \\
& + e^{-\frac{2\pi ij}{3}}\big( |2\,0,2\,0\rrangle^{P} +
 |2\,0,3\,1\rrangle^{P} \big) \ . 
\end{align}
Finally, the complete generalised permutation branes are then simply
given by  
\begin{equation}\label{final}
\begin{array}{rcl}
{\displaystyle |\!|P(j)\rrangle} 
& = & {\displaystyle |\!|P(j)\rrangle_{\text{NSNS}} + 
|\!|P(j)\rrangle_{\text{RR}}} \\
{\displaystyle \overline{|\!|P(j)\rrangle}}
& = & {\displaystyle
|\!|P(j)\rrangle_{\text{NSNS}} - 
|\!|P(j)\rrangle_{\text{RR}} \ , }
\end{array}
\end{equation}
where we have denoted anti-branes by a bar. We claim that 
this ansatz correctly reproduces the topological spectrum that is
predicted by the matrix factorisation calculation. This will be
checked next.

\subsection{Comparison of the spectra}

In order to compare with the results of section~2.1.1 we need to
calculate three different open string spectra, the self-overlap, the
relative overlap between the different $P$-branes, and the overlap with
the tensor product branes.

\subsubsection{The self-overlap}

The self-overlap of the NSNS contribution
$|\!|P(j)\rrangle_{\text{NSNS}}$ is in fact the same as that of the
$|\!|T(0,0)\rrangle_{\text{NSNS}}$ brane. It is easily seen to equal  
\begin{align}
_{\text{NSNS}}&\llangle P(j) |\!|
q^{\frac{1}{2} (L_{0}+\tilde{L}_{0})-\frac{c}{24}} |\!|
P(j)\rrangle_{\text{NSNS}}  \nonumber\\
= &\frac{1}{2} \big(\mmch[\tilde{q}]{1}{1}{1}+\mmch[\tilde{q}]{1}{3}{1}
+\mmch[\tilde{q}]{1}{1}{3} \big) \nonumber\\
&\times \big(\mmch[\tilde{q}]{4}{1}{1}
+\mmch[\tilde{q}]{4}{9}{1}+\mmch[\tilde{q}]{4}{7}{3}
+\mmch[\tilde{q}]{4}{5}{5}+\mmch[\tilde{q}]{4}{3}{7}
+\mmch[\tilde{q}]{4}{1}{9} \big) \ ,
\label{soTNS}
\end{align}
where $q=e^{2\pi i\tau }$ and $\tilde{q}=e^{-2\pi i/\tau }$.  The
spectrum factorises into the part coming from the $k_{1}=1$ factor and
into the part from the $k_{2}=4$ factor.
For the RR sector contribution we obtain on the other hand
\begin{align}
_{\text{RR}}\llangle P (j)|\!|& q^{\frac{1}{2}
(L_{0}+\tilde{L}_{0})-\frac{c}{24}} |\!|P (j)\rrangle_{\text{RR}}
\nonumber \\ 
&  \hspace*{-1.3cm} =   \frac{1}{2} \Big( \Rmmch[q]{1}{1}{1}
\Rmmch[q]{4}{1}{1}
+\Rmmch[q]{1}{1}{1}\Rmmch[q]{4}{2}{2}\Big) \nonumber\\
&  \hspace*{-1.3cm} \quad + \Big( \Rmmch[q]{1}{1}{0}\Rmmch[q]{4}{4}{0}
+\Rmmch[q]{1}{2}{0}\Rmmch[q]{4}{2}{0}+\Rmmch[q]{1}{2}{0}\Rmmch[q]{4}{3}{1}
\nonumber\\ 
&  \hspace*{-1.3cm} \qquad 
+\Rmmch[q]{1}{1}{0}\Rmmch[q]{4}{1}{3}-\Rmmch[q]{1}{1}{1} 
\Rmmch[q]{4}{1}{1}-\Rmmch[q]{1}{1}{1}\Rmmch[q]{4}{2}{2}\Big) \\
&  \hspace*{-1.3cm} =
\frac{1}{2}\bigg(\tmmch{1}{1}{1}\Big(\tmmch{4}{1}{1}
+\tmmch{4}{1}{9}
+\tmmch{4}{3}{7}-\tmmch{4}{5}{5}+\tmmch{4}{7}{3}+\tmmch{4}{9}{1}
\Big)\nonumber\\ 
&  \hspace*{-1.3cm} \quad 
+\tmmch{1}{3}{1}\Big(\tmmch{4}{1}{1}-\tmmch{4}{1}{9}
+\tmmch{4}{3}{7}
-\tmmch{4}{5}{5}-\tmmch{4}{7}{3}+\tmmch{4}{9}{1} \Big)\nonumber\\
& \hspace*{-1.3cm} \quad 
+\tmmch{1}{1}{3}\Big(\tmmch{4}{1}{1}+\tmmch{4}{1}{9}
-\tmmch{4}{3}{7}
-\tmmch{4}{5}{5}+\tmmch{4}{7}{3}-\tmmch{4}{9}{1} \Big)
 \bigg)\ . 
\label{spectrumPR}
\end{align}
Here $\tilde\chi$ denotes the character twisted by $(-1)^F$. It is
interesting to note that the spectrum can be written entirely in terms
of products of minimal model characters. 

The spectrum between the brane with itself is then the sum of these
two contributions; the  spectrum between brane and anti-brane is
obtained by subtracting the RR contribution from the NSNS one. In
either case it is clear from the above expressions that the resulting
spectrum satisfies the Cardy condition, {\it i.e.} that it consists of
an integer linear combination of GSO-projected $N=2$ representations.  

To compare these spectra with the results from the matrix
factorisation approach, we finally have to extract the topological
states. The chiral primaries in the spectrum are the ground states of
the representations $(u_1,v_1)\otimes (u_2,v_2)$ that have a
$v_1=v_2=\tfrac{1}{2}$. In the present case the relevant
representations are 
\begin{equation}
(\tfrac{1}{2},\tfrac{1}{2})\otimes (\tfrac{1}{2},\tfrac{1}{2})\ ,\
(\tfrac{3}{2},\tfrac{1}{2})\otimes (\tfrac{1}{2},\tfrac{1}{2})\ , \
(\tfrac{1}{2},\tfrac{1}{2})\otimes (\tfrac{9}{2},\tfrac{1}{2})\ ,\
(\tfrac{3}{2},\tfrac{1}{2})\otimes (\tfrac{9}{2},\tfrac{1}{2})  \ ,
\end{equation}
and their $U(1)$ charges are $q=0,\frac{1}{3},\frac{2}{3},1$,
respectively. In the brane-brane case all four states survive
(thus we have four `topological bosons'), while
for the brane anti-brane case all are projected away ({\it i.e.} there
are no `topological fermions'.)  This is then in perfect agreement
with the prediction~(1) in section~2.1.1.

\subsubsection[Relative overlaps between the $P$-branes]{Relative 
overlaps between the $\boldsymbol{P}$-branes}

In the NSNS sector the relative overlap of $P (j)$ with $P (j+1)$ is
\begin{align}
_{\text{NSNS}}\llangle P (j) |\!|& q^{\frac{1}{2} 
(L_{0}+\tilde{L}_{0})-\frac{c}{24}} |\!|P (j+1) \rrangle_{\text{NSNS}}
\nonumber\\
= &  
\frac{\sqrt{3}}{2} \Big( \mmch[q]{1}{1}{1}\mmch[q]{4}{1}{1} -\frac{4}{3} 
\sum_{m=1}^{\infty}\sum_{\substack{n=1\\
\makebox[0cm][c]{$\scriptstyle\mspace{-50mu}m\ \text{or}\ n\
  \text{odd}$}}}^{m} 
 \dgch[q]{m^{2}+n^{2}-mn}{0} \nonumber\\
& \quad + \mmch[q]{1}{1}{1}\mmch[q]{4}{5}{5}  -\frac{4}{3} 
\sum_{m=1}^{\infty}\sum_{\substack{n=1\\
\makebox[0cm][c]{$\scriptstyle\mspace{-50mu}m\ \text{or}\ n\
  \text{odd}$}}}^{m} 
 \dgch[q]{m^{2}+n^{2}-mn}{0} \nonumber\\
& \quad + 2\mmch[q]{1}{1}{1}\mmch[q]{4}{3}{3} -\frac{8}{3}
\sum_{m=0}^{\infty}\sum_{\substack{n=0\\ 
\makebox[0cm][c]{$\scriptstyle\mspace{-40mu}m+1\ \text{or}\ n\
\text{odd}$}}}^{\infty } \dgch[q]{\frac{1}{3}+m^{2}+n^{2}-mn+n}{0}
\Big)\nonumber \\
= & _{\text{NSNS}}\llangle P (j) 
|\!|q^{\frac{1}{2} (L_{0}+\tilde{L}_{0})-\frac{c}{24}} |\!|
P(j)\rrangle_{\text{NSNS}}  \ + \ \Pi (q) \ ,
\label{reloverlap}
\end{align}
where $\Pi $ denotes the difference to the self-overlap~\eqref{soTNS}
of the $P(j)$-brane. It is given by 
\begin{align}
\Pi (q) \ &=\ -\frac{4}{\sqrt{3}}\Big(
\sum_{m=1}^{\infty}\sum_{\substack{n=1\\ 
\makebox[0cm][c]{$\scriptstyle\mspace{-50mu}m\ \text{or}\ n\
  \text{odd}$}}}^{m} 
 \dgch[q]{m^{2}+n^{2}-mn}{0} + \sum_{m=0}^{\infty}\sum_{\substack{n=0\\ 
\makebox[0cm][c]{$\scriptstyle\mspace{-40mu}m+1\ \text{or}\ n\ 
\text{odd}$}}}^{\infty } \dgch[q]{\frac{1}{3}+m^{2}+n^{2}-mn+n}{0}
\Big)\nonumber\\ 
&= \  -\frac{4}{\sqrt{3}} \sum_{m=1}^{\infty}\sum_{\substack{n=1\\
\makebox[0cm][c]{$\scriptstyle\mspace{-50mu}m\ \text{or}\ n\
  \text{odd}$}}}^{m} 
 \dgch[q]{\frac{m^{2}+n^{2}-mn}{3}}{0}\nonumber\\
&=\  -\frac{2}{3\sqrt{3}} \Big( \sum_{m,n\in \mathbb{Z}}
q^{\frac{m^{2}+n^{2}-mn}{3}} - \sum_{\substack{m,n\in \mathbb{Z} \\
\makebox[0cm][c]{$\scriptstyle m,n\
\text{even}$}}}
q^{\frac{m^{2}+n^{2}-mn}{3}}\Big) \frac{\vartheta_{3} (q)}{\eta^{3} (q)}\ .
\end{align} 
Modular transformation of this term leads to
\begin{align}
\tilde{\Pi} (\tilde{q}) &=\ -\frac{2}{3\sqrt{3}} 
\Big( \sqrt{3} \sum_{m,n\in \mathbb{Z}}
\tilde{q}^{m^{2}+n^{2}-mn} -\frac{\sqrt{3}}{4} \sum_{m,n\in
\mathbb{Z}}\tilde{q}^{\frac{m^{2}+n^{2}-mn}{4}} \Big)
\frac{\vartheta_{3} (\tilde{q})}{\eta^{3} (\tilde{q})}\nonumber \\
&=\ -4
\Big(\sum_{m=1}^{\infty}\sum_{n=1}^{m}\tilde{q}^{m^{2}+n^{2}-mn}
-\frac{1}{4} \sum_{m=1}^{\infty}\sum_{n=1}^{m}
\tilde{q}^{\frac{m^{2}+n^{2}-mn}{4}} \Big) \frac{\vartheta_{3}
(\tilde{q})}{\eta^{3} (\tilde{q})} -\frac{1}{2}\frac{\vartheta_{3}
(\tilde{q})}{\eta^{3} (\tilde{q})}\nonumber \\
&=\ -3 \sum_{m=1}^{\infty}\sum_{n=1}^{m}
\dgch[\tilde{q}]{m^{2}+n^{2}-mn}{0} +
\sum_{m=1}^{\infty}\sum_{\substack{n=1\\ 
\makebox[0cm][c]{$\scriptstyle\mspace{-50mu}m\ \text{or}\ n\
\text{odd}$}}}^{m} \dgch[\tilde{q}]{\frac{m^{2}+n^{2}-mn}{4}}{0}
-\frac{1}{2}\frac{\vartheta_{3} (\tilde{q})}{\eta^{3} (\tilde{q})}
\nonumber \\
&=\ \sum_{m=1}^{\infty}\sum_{\substack{n=1\\
\makebox[0cm][c]{$\scriptstyle\mspace{-50mu}m\ \text{or}\ n\
\text{odd}$}}}^{m} \dgch[\tilde{q}]{\frac{m^{2}+n^{2}-mn}{4}}{0}
- \frac{1}{2} \mmch[\tilde{q}]{1}{1}{1}
\Big(\mmch[\tilde{q}]{4}{1}{1}+\mmch[\tilde{q}]{4}{5}{5}\Big)
\nonumber \\
&\qquad -\frac{1}{2} \mmch[\tilde{q}]{1}{3}{1}
\Big(\mmch[\tilde{q}]{4}{9}{1}+\mmch[\tilde{q}]{4}{3}{7}\Big) 
\nonumber\\
&\qquad  -\frac{1}{2} \mmch[\tilde{q}]{1}{1}{3}
\Big( \mmch[\tilde{q}]{4}{1}{9}+\mmch[\tilde{q}]{4}{7}{3} \Big)\ ,
\end{align}
where in the last step we used~\eqref{identitiese}. 
Inserting this and~\eqref{soTNS} in~\eqref{reloverlap} we finally
obtain 
\begin{align}
_{\text{NSNS}}\llangle P (j) |\!|& q^{\frac{1}{2}
(L_{0}+\tilde{L}_{0})-\frac{c}{24}} |\!| P(j+1) \rrangle_{\text{NSNS}}
\nonumber\\
= &
\frac{1}{2}\bigg(\mmch[\tilde{q}]{1}{1}{1}
\Big(\mmch[\tilde{q}]{4}{1}{9}+\mmch[\tilde{q}]{4}{3}{7}
+\mmch[\tilde{q}]{4}{7}{3}+\mmch[\tilde{q}]{4}{9}{1} \Big)\nonumber\\
& \quad +\mmch[\tilde{q}]{1}{3}{1}
\Big(\mmch[\tilde{q}]{4}{1}{1}+\mmch[\tilde{q}]{4}{1}{9}
+\mmch[\tilde{q}]{4}{5}{5}+\mmch[\tilde{q}]{4}{7}{3} \Big)\nonumber\\
&\quad +\mmch[\tilde{q}]{1}{1}{3}
\Big(\mmch[\tilde{q}]{4}{1}{1}+\mmch[\tilde{q}]{4}{3}{7}
+\mmch[\tilde{q}]{4}{5}{5}+\mmch[\tilde{q}]{4}{9}{1} \Big)
\bigg) \nonumber\\
&+ \sum_{m=1}^{\infty}\sum_{\substack{n=1\\
\makebox[0cm][c]{$\scriptstyle\mspace{-50mu}m\ 
\text{or}\ n\ \text{odd}$}}}^{m} 
\dgch[\tilde{q}]{\frac{m^{2}+n^{2}-mn}{4}}{0}\ .
\label{rspectrumTPNS}
\end{align}
Note that the spectrum can be described as the spectral flow orbits of
the two chiral primaries $(\frac{1}{2},\frac{1}{2})\otimes
(\frac{9}{2},\frac{1}{2})$ and $(\frac{3}{1},\frac{1}{2})\otimes
(\frac{1}{2},\frac{1}{2})$, and the tower of representations with
charge zero for which each representation is spectral flow invariant.

\noindent The overlap between different permutation branes in the RR
sector is 
\begin{align}
_{\text{RR}}\llangle P (j)|\!|&q^{\frac{1}{2}
(L_{0}+\tilde{L}_{0})-\frac{c}{24}}|\!|P (j+1)\rrangle_{\text{RR}} 
\nonumber\\
=  & \frac{1}{2} \Big( \Rmmch[q]{1}{1}{1}\Rmmch[q]{4}{1}{1}
+\Rmmch[q]{1}{1}{1}\Rmmch[q]{4}{2}{2}\Big)\nonumber\\
& -\frac{1}{2} \Big( \Rmmch[q]{1}{1}{0}\Rmmch[q]{4}{4}{0}
+\Rmmch[q]{1}{2}{0}\Rmmch[q]{4}{2}{0}+\Rmmch[q]{1}{2}{0}\Rmmch[q]{4}{3}{1}
\nonumber\\
& \qquad +\Rmmch[q]{1}{1}{0}\Rmmch[q]{4}{1}{3}-\Rmmch[q]{1}{1}{1}
\Rmmch[q]{4}{1}{1}-\Rmmch[q]{1}{1}{1}\Rmmch[q]{4}{2}{2}\Big)\\
=\ &-\frac{1}{2}\bigg(\tmmch{1}{1}{1}\Big(\tmmch{4}{1}{9}
+\tmmch{4}{3}{7}+\tmmch{4}{7}{3}+\tmmch{4}{9}{1} \Big)\nonumber\\
& \quad +\tmmch{1}{3}{1}\Big(\tmmch{4}{1}{1}-\tmmch{4}{1}{9}
-\tmmch{4}{5}{5}-\tmmch{4}{7}{3} \Big)\nonumber\\
&\quad +\tmmch{1}{1}{3}\Big(\tmmch{4}{1}{1}-\tmmch{4}{3}{7}
-\tmmch{4}{5}{5}-\tmmch{4}{9}{1} \Big)
 \bigg)\ .
\label{rspectrumPPR}
\end{align}
Combining these two expressions we thus see that the sum ({\it i.e.}
the overlap between brane and brane) does not contain any chiral
primaries. On the other hand the relative spectrum of $P (j)$ and
$\overline{P (j+1)}$ contains one  chiral primary in the sector
$(\frac{1}{2},\frac{1}{2})\otimes 
(\frac{9}{2},\frac{1}{2})$ with charge $q=\frac{2}{3}$ and a second in
the sector $(\frac{3}{2},\frac{1}{2})\otimes
(\frac{1}{2},\frac{1}{2})$ with charge $q=\frac{1}{3}$. This coincides
with the topological spectrum in prediction~(2) in section~2.1.1.

\subsubsection[The relative overlap between $P$-branes and the
$T(0,0)$-brane]{The relative overlap between $\boldsymbol{P}$-branes and the
$\boldsymbol{T(0,0)}$-brane} 

Finally, we consider the relative overlap between $P(j)$ and
$T(0,0)$. It is very easy to see that it is independent of $j$. In the
NSNS sector the result is actually the same as for the relative
overlap between $P(j)$ and $P(j+1)$ (\ref{rspectrumTPNS}). In the RR
sector we find on the other hand
\begin{align}
_{\text{RR}}& \llangle T (0,0) |\!| q^{\frac{1}{2}
(L_{0}+\tilde{L}_{0})-\frac{c}{24}}|\!| P (j)\rrangle_{\text{RR}}
\nonumber\\
=\ & - \frac{i\sqrt{3}}{2} \sum_{n=0}^{\infty}
\Big(\dgch[q]{\frac{1}{8}+1+3n}{\frac{1}{2}}
-\dgch[q]{\frac{1}{8}+1+3n}{-\frac{1}{2}}
-\dgch[q]{\frac{1}{8}+2+3n}{\frac{1}{2}} 
+\dgch[q]{\frac{1}{8}+2+3n}{-\frac{1}{2}}\nonumber \\
& \qquad \qquad -\dgch[q]{\frac{5}{8}+3n}{\frac{1}{2}} 
+\dgch[q]{\frac{5}{8}+3n}{-\frac{1}{2}} 
+\dgch[q]{\frac{5}{8}+2+3n}{\frac{1}{2}} 
-\dgch[q]{\frac{5}{8}+2+3n}{-\frac{1}{2}}  \Big)\\
=\ & - \frac{i\sqrt{3}}{2}\Big(-\Rmmch[q]{1}{2}{0} \Rmmch[q]{4}{2}{0}
+\Rmmch[q]{1}{1}{0}\Rmmch[q]{4}{4}{0} \nonumber \\
& \qquad - \Rmmch[q]{1}{2}{0}\Rmmch[q]{4}{3}{1} 
+\Rmmch[q]{1}{1}{0}\Rmmch[q]{4}{1}{3}\Big)\\
=\ & + \frac{1}{2}\tmmch{1}{1}{3} \Big( \tmmch{4}{1}{1} +
\tmmch{4}{9}{1} - \tmmch{4}{5}{5} + \tmmch{4}{3}{7} \Big) \nonumber\\
& - \frac{1}{2} \tmmch{1}{3}{1} 
\Big( \tmmch{4}{1}{1} + \tmmch{4}{1}{9}
- \tmmch{4}{5}{5} + \tmmch{4}{7}{3} \Big)\nonumber\\
& - \frac{1}{2}\tmmch{1}{1}{1} \Big( \tmmch{4}{1}{9} + \tmmch{4}{7}{3}
- \tmmch{4}{3}{7} - \tmmch{4}{9}{1} \Big)\ .
\label{rspectrumTPR}
\end{align}
Here we used the character decompositions of appendix~B.2 in the first step.
By adding the two contributions (eqs.\ \eqref{rspectrumTPNS}
and~\eqref{rspectrumTPR}) we thus find that the 
relative spectrum between $T (0,0)$ and $P (j)$ 
contains a chiral primary in the sector 
$(\frac{1}{2},\frac{1}{2})\otimes
(\frac{9}{2},\frac{1}{2})$ of charge $q=\frac{2}{3}$. 
The relative spectrum between $T (0,0)$ and $\overline{P (j)}$ on the
other hand contains one chiral primary in the sector
$(\frac{3}{2},\frac{1}{2})\otimes (\frac{1}{2},\frac{1}{2})$ of charge
$q=\frac{1}{3}$. This is then again consistent with the prediction~(3)
of section~2.1.1.
\bigskip 

The correct topological spectrum is a strong indication that the
branes we have constructed above correspond indeed to the generalised
permutation factorisations. We shall now give further evidence for the
consistency of the $P$-branes by relating them to constructions in a
free orbifold theory.

\section{The torus orbifold} 

The $A_{1}\times A_{4}$ we discussed in the previous section has
$c=3$, and one may therefore expect that it is in fact equivalent to a
free field theory. This is indeed the case: the theory is 
equivalent to a $\mathbb{Z}_{6}$-orbifold of the $su(3)$ torus
\cite{Lerche:1989cs}. The boundary states we have constructed above
should thus have a description in terms of this free field theory;
this will now be explained.

\subsection{The orbifold theory}

We consider the $N=2$ theory of two bosons and two fermions on the 
$su(3)$ torus. The torus is described by the two-dimensional plane
with identifications given by two vectors, $e_1,e_2$ where 
$e_1 \cdot e_1 = e_2 \cdot e_2 = 2$ and $e_1 \cdot e_2=-1$. If we
think of the plane as $\mathbb{C}$, then we can choose $e_1,e_2$ as two
vectors of equal length with $e_1$ pointing along the positive real
axis, and $e_2$ pointing in the direction of $e^{\frac{2\pi i}{3}}$.
The winding on the torus is then described by the lattice
$\Lambda_{R}$ spanned by $e_1$ and $e_2$. The momentum takes values in
the dual lattice $\Lambda_{W}$ spanned by 
\begin{equation}
e_1^*= \frac{2}{3} e_1 + \frac{1}{3} e_2 \quad , \quad 
e_2^*= \frac{1}{3} e_1 + \frac{2}{3} e_2 
\end{equation}
with the inner products 
$e_1^* \cdot e_1^*=e_2^*\cdot e_2^*=\frac{2}{3}$ and 
$e_1^* \cdot e_2^* = \frac{1}{3}$. The lattices $\Lambda_{R}$ and
$\Lambda_{W}$ are the root and weight lattice of $su (3)$,
respectively. 

\noindent The theory that is of interest to us has in addition a
B-field, and the left- and right-moving momenta are thus given by 
\begin{equation}\label{leftrightmomenta}
p_L = p+\tfrac{1}{2}L -\tfrac{1}{2}B L \ , \qquad 
p_R = p-\tfrac{1}{2}L  -\tfrac{1}{2}B L\ , 
\end{equation} 
where $p\in \Lambda_{W}$ and $L\in \Lambda_{R}$. The B-field is a
matrix chosen such that $B e_{1} = -e_{2}^{*}$ and $B
e_{2}=e_{1}^{*}$. Then the lattice $\Gamma_{2,2}$ of the momenta
$(p_{L},p_{R})$ is
\begin{equation}
\Gamma_{2,2} = \{(p_{L},p_{R})\in \Lambda_{W}\oplus \Lambda_{W}\ ,\
p_{L}-p_{R}\in \Lambda_{R}  \}\ .
\end{equation}
The resulting partition function is (we are considering a type 0B like
GSO-projection) 
\begin{equation}
Z_{\text{Torus}} = \frac{1}{|\eta(q)|^4} \sum_{(p_{L},p_{R})\in\Gamma_{2,2}}
q^{\frac{1}{2}(p_{L})^2}  
\bar{q}^{\frac{1}{2}(p_{R})^2} 
\bigg( \frac{|\vartheta_3(q)|^2}{|\eta(q)|^2} + 
\frac{|\vartheta_2(q)|^2}{|\eta(q)|^2} +
\frac{|\vartheta_4(q)|^2}{|\eta(q)|^2}  \bigg) \ .
\end{equation}
Our conventions for the $\vartheta$ and $\eta$ functions are summarised
in appendix~B.4.
In the following we shall use the left-moving modes $\alpha_n$ and
$\bar\alpha_n$ corresponding to complex target space coordinates; the 
right-moving modes will be denoted by a tilde. Similarly, we use
complex fermionic oscillator modes  $\psi^+_r$ and $\psi^-_r$. The
commutation relations are then
\begin{align}
[\alpha_{m},\bar{\alpha}_{n}] = & m\, \delta_{m,-n} \ , & 
\{\psi_{r}^{+} , \psi_{s}^{-} \} = & \delta_{r,-s} \ ,
\end{align}  
with similar relations for the right-movers. The
$\mathbb{Z}_{6}$-orbifold of interest acts on the complex coordinates
as $z\to e^{2\pi i/6} z$. It is not difficult to show that the
spectrum of this orbifold theory agrees indeed with (\ref{k4k1spec}). 

\subsection[Branes on the $su(3)$ torus]{Branes on the 
$\boldsymbol{su(3)}$ torus} 

The branes we are interested in are D1-branes on the torus with Wilson
lines.\footnote{This is similar to the situation in
\cite{Dell'Aquila:2005jg}, see also \cite{Brunner:2006tc}.}
The relevant gluing conditions are
\begin{align}
(\alpha_n + e^{2i\phi} \tilde{\bar{\alpha}}_{-n})\,
|\!|B\rrangle &= 0 & 
(\bar{\alpha}_n + e^{-2i\phi} \tilde{\alpha}_{-n})\,
|\!|B\rrangle &=  0 \\
(\psi^+_r - ie^{2i\phi} \eta \tilde{\psi}^-_{-r})\,
|\!|B\rrangle  &=  0 &
(\psi^-_r +ie^{-2i\phi} \eta\tilde{\psi}^+_{-r})\,
|\!|B\rrangle  &=  0\ .
\end{align}
The boundary states are linear combinations of the coherent states 
\begin{align}
|\phi,p_L,\eta \rrangle = \exp \Big( & -\sum_{n=1}^{\infty} 
\frac{1}{n}\big( e^{-2i\phi} \alpha_{-n} \tilde{\alpha}_{-n} 
+ e^{2i\phi} \bar{\alpha}_{-n} \tilde{\bar{\alpha}}_{-n} \big)\\
&-i\eta \sum_{r>0} \big( e^{-2i\phi} \psi^+_{-r} \tilde{\psi}^+_{-r}
-e^{2i\phi} \psi^-_{-r} \tilde{\psi}^-_{-r} \big)
\Big) |p_L,p_R=-e^{2i\phi}\bar{p}_L \rangle \ ,
\end{align}
where $|p_{L},p_{R}\rangle$ denotes the ground state with the
corresponding momenta. Note that the GSO projection (type 0B) is
chosen such that the boundary states of the torus theory (before
orbifolding) will not have any RR contribution. As before, we only
want to consider branes with a fixed spin-structure $\eta$, say
$\eta=+1$.  

\noindent Before orbifolding, the full boundary state is then
\begin{equation}\label{torusboundarystate}
|\!|\tilde{B} (u,\phi ) \rrangle \ =\ \mathcal{N} _{\phi}\sum_{p_L}
e^{2\pi i(u,p_L)} |\phi ,p_L \rrangle \ .
\end{equation}
Here $u$ is a vector which contains information on the position and
the Wilson line on the brane. The sum over $p_L$ is restricted to
those $p_L$ for which $p_{L}$ and $p_{R}=-e^{2i\phi}\bar{p}_{L}$ are 
allowed combinations of momenta. In particular, because of 
\eqref{leftrightmomenta}, the winding $L$ has to be in the direction
given by $e^{i\phi}$ (the direction of the brane), and the momentum
$p$ is perpendicular to the brane. The phase in front of the Ishibashi 
states in~\eqref{torusboundarystate} can be written as
\begin{equation}
e^{2\pi i (u,p_L)} = e^{2\pi i [u_{\perp}p 
+ \frac{1}{2}(u+B u)_\parallel L]} \ , 
\end{equation}
where $u_\parallel$ denotes the projection of $u$ in the direction of
the brane, and $u_\perp$ is the orthogonal complement. The boundary
state~\eqref{torusboundarystate} thus corresponds to a D1-brane in a
direction given by $e^{i\phi}$ in the complex plane with a position
given by $u_\perp$ and a Wilson line determined by 
$\frac{1}{2}(u+B u)_\parallel$. Note that the value of $u$ is a
priori only defined modulo an overall shift which corresponds to a
redefinition of the Ishibashi states by a phase. 

There are two types of angles that will be relevant in the
following. If $\phi_{n}=\frac{2\pi n}{3}$, then all momenta 
$p\in \Lambda_{W}$ are allowed in the sum
in~\eqref{torusboundarystate}. In this case the correctly normalised
boundary state is  
\begin{equation}
|\!|\tilde{B} (u,\phi_{n} ) \rrangle \ =\ 3^{-\frac{1}{4}} \ 
\sum_{p_L\in\Lambda_W}
e^{2\pi i(u,p_L)} |\phi_{n} ,p_L \rrangle 
\ . 
\end{equation}
Its self-spectrum is 
\begin{equation}
\llangle \tilde{B} (u,\phi_{n} )|\!| q^{\frac{1}{2}
(L_{0}+\tilde{L}_{0})-\frac{c}{24}} |\!| \tilde{B} (u,\phi_{n} )\rrangle
= \frac{1}{\sqrt{3}} \  \sum_{p\in \Lambda_{W}}
q^{\frac{1}{2}p^{2}}\frac{\vartheta_{3}(q)}{\eta^{3} (q)} 
 =  \sum_{p\in \Lambda_{R}}
\tilde{q}^{\frac{1}{2}p^{2}} \frac{\vartheta_{3} (\tilde{q})}{\eta^{3}
(\tilde{q})} \ . \label{ssbeforeorbifold}
\end{equation}
Here $\vartheta_{3} (q)$ denotes a Jacobi theta function, and $\eta
(q)$ is the Dedekind function. We also note that the 
relative spectrum between two branes 
$|\!|\tilde{B} (u,\phi )\rrangle$ with $\phi_{n}=\frac{2\pi n}{3}$ but
different labels $u,u'$ is
\begin{align}
\llangle \tilde{B} (u',\phi_{n} ) |\! |q^{\frac{1}{2}
(L_{0}+\tilde{L}_{0})-\frac{c}{24}} |\!| 
\tilde{B} (u,\phi_{n} )\rrangle =
&\frac{1}{\sqrt{3}} \sum_{p\in \Lambda_{W}}
q^{\frac{1}{2}p^{2}}e^{2\pi i(u-u',p)}
\frac{\vartheta_{3}(q)}{\eta^{3} (q)} \\
 = & \sum_{p\in \Lambda_{R}}
\tilde{q}^{\frac{1}{2} (p+u-u')^{2}} \frac{\vartheta_{3} 
(\tilde{q})}{\eta^{3} 
(\tilde{q})} \ .
\end{align}
Finally, the relative spectrum between branes at different angles
$\phi_{n}$ and $\phi_{n'}$ is ($n\not= n'$)
\begin{align}
\llangle \tilde{B} (u',\phi_{n'} )|\!| q^{\frac{1}{2}
(L_{0}+\tilde{L}_{0})-\frac{c}{24}} |\!| \tilde{B} (u,\phi_{n} )\rrangle 
= &  \frac{\vartheta_{3} (e^{2\pi i/3},q)}{\vartheta_{1} (e^{2\pi
i/3},q)}
=  -i \frac{\vartheta_{3} (\tilde{q}^{-1/3},\tilde{q})}{\vartheta_{1}
(\tilde{q}^{-1/3},\tilde{q})} 
\ .
\label{rspectrumdiffphi}
\end{align}

\noindent The other case of interest is 
$\phi=\psi_n=2\pi(\frac{1}{12}+\frac{n}{6})$ with $n=0,1,2$. The
corresponding boundary states then only couple to momenta $(p_L,p_R)$
from the root lattice, and the correctly normalised boundary state is 
\begin{equation}\label{torusboundarystate2}
|\!|\tilde{B} (u,\psi_n ) \rrangle \ =\ 3^{\frac{1}{4}} 
\sum_{p_L \in \Lambda_R} e^{2\pi i(u,p_L)} |\psi_n ,p_L \rrangle 
\end{equation}
with the self-spectrum
\begin{equation}
\llangle \tilde{B} (u,\psi_{n})|\!|q^{\frac{1}{2}
(L_{0}+\tilde{L}_{0})-\frac{c}{24}} |\!| \tilde{B} (u,\psi_{n} )\rrangle
=  \sum_{p\in \Lambda_{W}}
\tilde{q}^{\frac{1}{2}p^{2}} \frac{\vartheta_{3} (\tilde{q})}{\eta^{3}
(\tilde{q})} \ .
\end{equation}

\subsection{Branes on the orbifold}

It is convenient to perform the $\mathbb{Z}_{6}$-orbifold operation in
two steps, first by performing the $\mathbb{Z}_{2}$-part
(reflection). In a second step we then perform the remaining
$\mathbb{Z}_{3}$-orbifold (rotation by $120$ degrees). In the first
step there are the four $\mathbb{Z}_{2}$-fixed points (see
figure~\ref{fig:fixedpoints}). For each of the four fixed points we
have a $\mathbb{Z}_2$-twisted sector. The second
$\mathbb{Z}_3$-operation leaves the fixed point (0) 
invariant,  and permutes the three fixed points 
$(1) \mapsto (2) \mapsto (3)$. In addition to (0) it has two further
fixed points that lead again to twisted sectors. Some of the branes
are invariant under the reflection, and thus couple to
$\mathbb{Z}_2$-twisted sectors; however, none will be fixed under the
second $\mathbb{Z}_3$-orbifold, and we will therefore not have to
resolve any fixed points in the second step. 

\FIGURE{\begin{picture}(0,0)%
\includegraphics{fixedpoints.pstex}%
\end{picture}%
\setlength{\unitlength}{4144sp}%
\begingroup\makeatletter\ifx\SetFigFont\undefined%
\gdef\SetFigFont#1#2#3#4#5{%
  \reset@font\fontsize{#1}{#2pt}%
  \fontfamily{#3}\fontseries{#4}\fontshape{#5}%
  \selectfont}%
\fi\endgroup%
\begin{picture}(2556,1743)(2881,-2704)
\put(3691,-2626){\makebox(0,0)[lb]{\smash{{\SetFigFont{12}{14.4}{\rmdefault}{\mddefault}{\updefault}{$(0)$}%
}}}}
\put(4501,-2626){\makebox(0,0)[lb]{\smash{{\SetFigFont{12}{14.4}{\rmdefault}{\mddefault}{\updefault}{$(1)$}%
}}}}
\put(3016,-1681){\makebox(0,0)[lb]{\smash{{\SetFigFont{12}{14.4}{\rmdefault}{\mddefault}{\updefault}{$(2)$}%
}}}}
\put(3871,-1681){\makebox(0,0)[lb]{\smash{{\SetFigFont{12}{14.4}{\rmdefault}{\mddefault}{\updefault}{$(3)$}%
}}}}
\end{picture}%

\caption{\label{fig:fixedpoints}Fixed points}} 

Without loss of generality we may choose the phase factor of the
above Ishibashi states so that they transform trivially under a 
a rotation by $e^{2\pi i/6}$ 
\begin{equation}
|\phi,p \rrangle \to
|\phi+\tfrac{2\pi}{6},e^{2\pi i/6}p \rrangle  \ .
\end{equation}

\subsubsection[Orbifolding by the $\mathbb{Z}_2$-subgroup]{Orbifolding
by the $\boldsymbol{\mathbb{Z}_2}$-subgroup}

The D1-brane is fixed under the $\mathbb{Z}_{2}$-action if $u$ 
is of the form (written in the basis $(e_{1},e_{2})$) 
\begin{align}
u_{0}= & (0,0) & u_{1}= & (\tfrac{1}{2},0) \\
u_{2}= & (0,\tfrac{1}{2}) & u_{3}= & (\tfrac{1}{2},\tfrac{1}{2}) \ . 
\end{align}
If $u$ is of this form, the corresponding brane couples to some linear
combination of the four $\mathbb{Z}_2$-twisted sectors. With our
choice of GSO projection we do not get any Ishibashi  
states from the twisted NSNS sector, but we do obtain twisted RR
contributions which we denote by $| (i),\phi \rrangle$; we chose the
convention that Ishibashi states corresponding 
to different angles $\phi$ have the same phase in front of the ground 
state. Hence their overlaps are 
\begin{equation}
\llangle (i),\phi '|q^{\frac{1}{2}
(L_{0}+\tilde{L}_{0})-\frac{c}{24}}z^{\frac{1}{2} (J_{0}+\tilde{J}_{0})} 
| (i),\phi \rrangle =
 \frac{\vartheta_{3} (ze^{2i (\phi'-\phi)},q)}{\vartheta_{4} 
(e^{2i (\phi'-\phi)},q)} 
=  e^{-\frac{i\pi w^{2}}{\tau}} \tilde{z}^{\frac{\phi-\phi'}{\pi}}
\frac{\vartheta_{3}
(\tilde{z}\tilde{q}^{\frac{\phi-\phi'}{\pi}},\tilde{q})}{\vartheta_{2}
(\tilde{q}^{\frac{\phi-\phi'}{\pi}},\tilde{q})} \ .
\end{equation}
We included the $U(1)$ charges which we shall need later, and
introduced the notation $z=e^{2\pi iw}$ and $\tilde{z}=e^{2\pi iw/\tau }$. 

Depending on the Wilson line and the orientation, the different
D1-branes couple differently to the twisted sectors. 
For $\phi=0$, we have formally the structure ($(i)$ denotes the 
contribution of the $i^{\text{th}}$ twisted sector) 
\begin{subequations}\label{twistedsectors}
\begin{align}
u_{0}:\ & (0) + (1)  & u_{1}:\ & (1) - (0) \\
u_{2}:\ & -(2) - (3)  & u_{3}:\ & (2) - (3) \ .
\end{align}
Analogously we then have for $\phi =2\pi /3$:
\begin{align}
u_{0}:\ & (0) + (2)  & u_{2}:\ & (2) - (0) \\
u_{3}:\ & -(3) - (1)  & u_{1}:\ & (3) - (1) \ ,
\end{align}
and for $\phi =4\pi /3$:
\begin{align}
u_{0}:\ & (0) + (3)  & u_{3}:\ & (3) - (0) \\
u_{1}:\ & -(1) - (2)  & u_{2}:\ & (1) - (2) \ .
\end{align}
\end{subequations}
A boundary state in the $\mathbb{Z}_{2}$-orbifold is then
\begin{equation}
|\!|B (u_{i},\phi_{n})\rrangle = 
\frac{1}{\sqrt{2}} |\!|\tilde{B} (u_i,\phi_{n})\rrangle
+ \frac{1}{\sqrt{2}}\big(\text{twisted sector contribution}\big) \ .
\end{equation}
The corresponding anti-brane has the opposite sign in front of the 
twisted sector contribution. The spectrum of the resolved branes is
then given by
\begin{align}
\llangle B (u_{i},\phi_{n})|\!| q^{\frac{1}{2}
(L_{0}+\tilde{L}_{0})-\frac{c}{24}} |\!| B (u_{i},\phi_{n})\rrangle
= & \frac{1}{2} \sum_{p\in \Lambda_{R}}
\tilde{q}^{\frac{1}{2}p^{2}} \frac{\vartheta_{3} (\tilde{q})}{\eta^{3}
(\tilde{q})} + \frac{\vartheta_{3}
(\tilde{q})}{\vartheta_{2}
(\tilde{q})} \ .
\end{align} 
\medskip

With these preparations we can now specify
all the relative spectra between the resolved branes. If the angles
are equal, but the Wilson line parameters $u_{i},u_{i'}$ are
different, then there is no contribution from the twisted sector and
we find  
\begin{align}
\llangle B (u_{i}',\phi_{n})|\!| q^{\frac{1}{2}
(L_{0}+\tilde{L}_{0})-\frac{c}{24}} |\!|B (u_{i},\phi_{n})\rrangle
= & \frac{1}{2} \sum_{p\in \Lambda_{R}}
\tilde{q}^{\frac{1}{2} (p+u_{1})^{2}} 
\frac{\vartheta_{3} (\tilde{q})}{\eta^{3}
(\tilde{q})}
\ .
\end{align} 
Here we have used that $u_{i'}-u_{i}=u_{1}\mod \Lambda_{R}$ for
$i\not= i'$. 

If the angles are different, there is always a contribution from the 
twisted sector, and the spectrum is given by
\begin{equation}\label{twsectorcontribution}
\frac{1}{2}
\Big(-i\frac{\vartheta_{3}(\tilde{q}^{-1/3},\tilde{q})}{\vartheta_{1}
(\tilde{q}^{-1/3},\tilde{q})} \pm \frac{\vartheta_{3}
(\tilde{q}^{-1/3},\tilde{q})}{\vartheta_{2}
(\tilde{q}^{-1/3},\tilde{q})} \Big) \ .
\end{equation}
The sign depends on the couplings to the twisted sectors; this can be 
determined from \eqref{twistedsectors}.

\subsubsection[The final $\mathbb{Z}_3$-orbifold]{The 
final $\boldsymbol{\mathbb{Z}_3}$-orbifold}

We can now take superpositions of these D-branes under the remaining
$\mathbb{Z}_{3}$-action to turn them into branes of the 
$\mathbb{Z}_6$-orbifold. In particular, we find the four branes that
we label as  
\begin{align}
|\!|T (0,0)\rrangle = & \frac{1}{\sqrt{3}} 
\big(|\!|B (u_{3},\phi =0)\rrangle +
|\!|B (u_{1},\phi =\tfrac{2\pi}{3})\rrangle  
+ |\!|B (u_{2},\phi =\tfrac{4\pi}{3})\rrangle \big)\\
|\!|P (1)\rrangle = & \frac{1}{\sqrt{3}} \big(
|\!|B (u_{0},\phi =0)\rrangle  +
|\!|B (u_{0},\phi =\tfrac{2\pi}{3})\rrangle  
+ |\!|B (u_{0},\phi =\tfrac{4\pi}{3})\rrangle \big)\\
|\!|P (2)\rrangle = & \frac{1}{\sqrt{3}} \big(
|\!|B (u_{1},\phi =0)\rrangle  +
|\!|B (u_{2},\phi =\tfrac{2\pi}{3})\rrangle  
+ |\!|B (u_{3},\phi =\tfrac{4\pi}{3} )\rrangle \big)\\
|\!|P (3)\rrangle = & \frac{1}{\sqrt{3}} \big(
|\!|B (u_{2},\phi =0)\rrangle  +
|\!|B (u_{3},\phi =\tfrac{2\pi}{3})\rrangle  
+ |\!|B (u_{1},\phi =\tfrac{4\pi}{3})\rrangle \big) \ .
\end{align}
The corresponding anti-branes differ in their coupling to the twisted
RR sectors. Figures~\ref{fig:t1} and~\ref{fig:t2} show an illustration
of these branes on the torus. Note that the branes $P (1)$ and $P (2)$
($T (0,0)$ and $P (3)$) are at the same location, but carry different
Wilson lines which is not visible in the figure.
\DOUBLEFIGURE{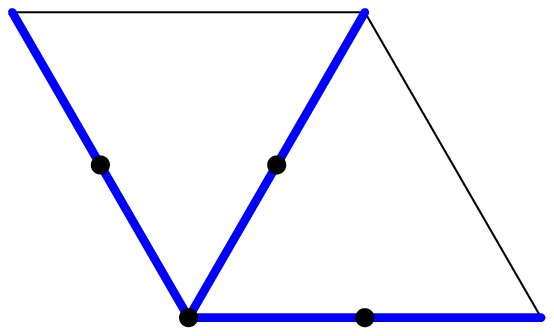}{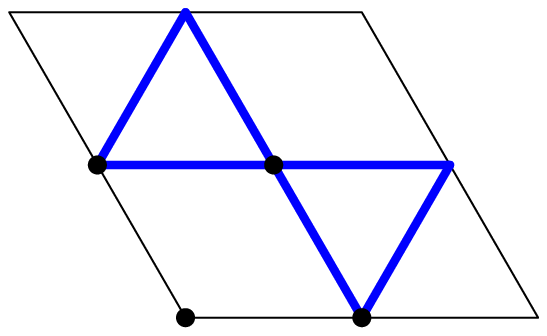}{\label{fig:t1}The superposition of
branes that make up $P(1)$ or $P (2)$.}{\label{fig:t2}The 
superposition of branes that make up $T (0,0)$ or $P(3)$.}

We now claim that these branes agree indeed with the correspondingly
labelled branes of the $N=2$ theory. In the following we shall confirm
this by determining the relative spectra and comparing them with the
results of the previous section.

\subsection{Comparison of the overlaps}

It is straightforward to determine the self-spectra of these orbifold
branes. They are 
\begin{align}
\llangle T (0,0)|\!| q^{\frac{1}{2} (L_{0}+\tilde{L}_{0})-\frac{c}{24}}
|\!| T (0,0)\rrangle = & \frac{1}{2} \sum_{p\in \Lambda_{R}} 
\tilde{q}^{\frac{1}{2}p^{2}} \frac{\vartheta_{3} (\tilde{q})}{\eta^{3}
(\tilde{q})} + \frac{\vartheta_{3}
(\tilde{q})}{\vartheta_{2}
(\tilde{q})} \nonumber\\
 & + \bigg(-i\frac{\vartheta_{3}(\tilde{q}^{-1/3},\tilde{q})}{\vartheta_{1}
(\tilde{q}^{-1/3},\tilde{q})} - \frac{\vartheta_{3}
(\tilde{q}^{-1/3},\tilde{q})}{\vartheta_{2}
(\tilde{q}^{-1/3},\tilde{q})} \bigg) 
\end{align}
and
\begin{align}
\llangle P (j) |\!| 
q^{\frac{1}{2} (L_{0}+\tilde{L}_{0})-\frac{c}{24}}|\!|
P (j)\rrangle = & \frac{1}{2} \sum_{p\in \Lambda_{R}}
\tilde{q}^{\frac{1}{2}p^{2}} \frac{\vartheta_{3} (\tilde{q})}{\eta^{3}
(\tilde{q})} + \frac{\vartheta_{3}
(\tilde{q})}{\vartheta_{2}
(\tilde{q})} \nonumber\\
 & + \bigg(-i\frac{\vartheta_{3}(\tilde{q}^{-1/3},\tilde{q})}{\vartheta_{1}
(\tilde{q}^{-1/3},\tilde{q})} + \frac{\vartheta_{3}
(\tilde{q}^{-1/3},\tilde{q})}{\vartheta_{2}
(\tilde{q}^{-1/3},\tilde{q})} \bigg) \ .
\end{align}
The only difference in the spectra of the $T(0,0)$-brane and the 
$P (j)$-branes is the sign in the second line. 

\noindent The relative spectrum between the different $P (j)$-branes
is given by 
\begin{align}
\llangle P (j) |\!|
q^{\frac{1}{2} (L_{0}+\tilde{L}_{0})-\frac{c}{24}}|\!| P (j+1)\rrangle 
= & \frac{1}{2} \sum_{p\in \Lambda_{R}}
\tilde{q}^{\frac{1}{2} (p+u_{1})^{2}} 
\frac{\vartheta_{3} (\tilde{q})}{\eta^{3}
(\tilde{q})}\nonumber\\ 
& + \bigg(-i\frac{\vartheta_{3}(\tilde{q}^{-1/3},\tilde{q})}
{\vartheta_{1}(\tilde{q}^{-1/3},\tilde{q})} - \frac{\vartheta_{3}
(\tilde{q}^{-1/3},\tilde{q})}{\vartheta_{2}
(\tilde{q}^{-1/3},\tilde{q})} \bigg) \ .
\end{align}
For the relative spectrum between $T (0,0)$ and $P (j)$ we should
keep track of the $U (1)$-charges in the spectrum since otherwise the
RR contributions cancel in the overlap. We obtain
\begin{align}
\llangle T (0,0)|\!| &q^{\frac{1}{2}
(L_{0}+\tilde{L}_{0})-\frac{c}{24}}z^{\frac{1}{2} (J_{0}+\tilde{J}_{0})} 
|\!| P (j)\rrangle \  e^{\frac{i\pi w^{2}}{\tau } }\nonumber\\
= & 
\frac{1}{2} \sum_{p\in \Lambda_{R}}
\tilde{q}^{\frac{1}{2} (p+u_{1})^{2}} 
\frac{\vartheta_{3} (\tilde{z},\tilde{q})}{\eta^{3}
(\tilde{q})} \nonumber\\
& + \frac{1}{2}
\bigg(-i\tilde{z}^{-\frac{1}{3}} 
\frac{\vartheta_{3}(\tilde{z}\tilde{q}^{-1/3},\tilde{q})}{\vartheta_{1}
(\tilde{q}^{-1/3},\tilde{q})} -i\tilde{z}^{\frac{1}{3}}
\frac{\vartheta_{3}(\tilde{z}^{-1}\tilde{q}^{-1/3},\tilde{q})}
{\vartheta_{1}(\tilde{q}^{-1/3},\tilde{q})} \nonumber\\
& \qquad \quad + \tilde{z}^{-\frac{1}{3}} 
\frac{\vartheta_{3}(\tilde{z}\tilde{q}^{-1/3},\tilde{q})}{\vartheta_{2}
(\tilde{q}^{-1/3},\tilde{q})} 
-\tilde{z}^{\frac{1}{3}}
\frac{\vartheta_{3}(\tilde{z}^{-1}\tilde{q}^{-1/3},\tilde{q})}
{\vartheta_{2}(\tilde{q}^{-1/3},\tilde{q})} \bigg) \ .
\end{align}
Using the formulae of appendix~B.4, these spectra are easily seen to
reproduce those of section~2.3.
\medskip

Actually, one can also identify the other two tensor product branes 
($T(0,1)$ and $T(0,2)$) in the orbifold theory. The $T(0,1)$ brane 
corresponds to a superposition of branes at angles
$\psi_{n}$, perpendicular to the branes at angles $\phi_{n}$ which
made up the $T (0,0)$-brane. To see this one observes that the
relative angle to the $T (0,0)$-brane can be read off from the phases
in front of the Ishibashi states which come from the vacuum
sector of the torus. The $T(0,1)$-brane couples with a relative sign
to the sector 
$(\frac{1}{2},\frac{1}{2})\otimes (\frac{5}{2},\frac{5}{2})$, and 
thus in the relative overlap of the $T(0,0)$- and the $T(0,1)$-brane
the states with total conformal weight $H=1$ (and total charge $Q=0$)
do not contribute. This means that the $T(0,1)$-brane is perpendicular
to the $T(0,0)$ brane. 

By the same token it follows that the $T (0,2)$-brane is again a
superposition of branes at angles $\phi_{n}$. Its parameter $u$ cannot
be one of the four fixed values $u_{i}$ as it does not couple to the
twisted sector. A detailed analysis shows that for the $\phi =0$
direction the parameter $u$ is given by 
$u=u_{3}+ (\tfrac{1}{3},\tfrac{2}{3})$. 
\medskip

We have thus identified the various branes, including in particular
the generalised permutation branes $P(j)$, with standard constructions 
of the $T^2/\mathbb{Z}_6$ orbifold. This gives strong support to the
claim that these $P(j)$ branes are indeed consistent.

\subsection{The factorisation constraint}

Another consistency condition that can at least partially be checked
for these D-branes is the sewing constraint of 
\cite{Lewellen:1991tb} that is sometimes referred to as the 
factorisation constraint \cite{Recknagel:1998ih} or the `classifying
algebra' \cite{Fuchs:1997kt}. It requires that the coefficients in
front of the Ishibashi states of every consistent (fundamental)
boundary state satisfy a quadratic equation. (This equation comes from
considering different limits of the 2-point function of two bulk
fields in the presence of the boundary --- see for example
\cite{Gaberdiel:2001xm}.) In general this quadratic equation is
difficult to determine since it requires knowledge of the actual 
operator product expansion coefficients of the bulk theory, as well as
the fusing matrices. However, in our situation, at least some of these
relations can be easily found. 

Consider for example the (uncharged) primary fields of the diagonal
$N=2$ algebra that originate from the tensor product of the two NS
vacuum representations. As follows from the analysis of appendix~B,
in particular (\ref{vacdec1}),
these primary fields are labelled by $(m,n)$ where $m\geq n \geq 1$,
as well as the vacuum $(m,n)=(0,0)$. Since these fields arise in the 
`vacuum sector' their operator product expansion must close among
themselves, and they must therefore give rise to a factorisation
constraint. In fact, in terms of the $\mathbb{Z}_{6}$-orbifold
description, the primary field $(m,n)$ corresponds to the 
$\mathbb{Z}_{6}$-orbit that contains the momentum state corresponding
to $p=m e_1 + n e_2$. Given that the operator product expansion and
the fusing matrices of pure momentum states are very simple, it then
follows that the corresponding factorisation constraint is 
\begin{align}
B_{m_1,n_1} B_{m_2,n_2} = 
\frac{1}{6} \Big( &B_{m_1+m_2,n_1+n_2}+ B_{m_1-n_2,n_1
+m_2-n_2} + B_{m_1-m_2+n_2,n_1-m_2} \nonumber \\
+& B_{m_1-m_2,n_1-n_2} + B_{m_1+n_2,n_1-m_2+n_2} 
+ B_{m_1+m_2-n_2,n_1+m_2}\Big) \ . \label{facto}
\end{align}
Here $B_{m,n}$ is the coefficient of the $(m,n)$ Ishibashi state
from the $(\tfrac{1}{2}\tfrac{1}{2},\tfrac{1}{2}\tfrac{1}{2})$ sector
(normalised so that $B_{0,0}=1$). Furthermore, it is understood that
we have the identifications that come from the $\mathbb{Z}_{6}$-orbit 
\begin{equation}
B_{m,n} = B_{-n,m-n} = B_{n-m,-m} = B_{-m,-n} = B_{n,n-m}=B_{m-n,m} \
; \end{equation}
this allows us to rewrite every term on the right-hand-side in terms
of a coefficient corresponding to $(m,n)$ with $m\geq n \geq 1$
or $(m,n)=(0,0)$.

It is obvious that the original tensor branes (for which $B_{m,n}=1$
for all $(m,n)$) satisfy (\ref{facto}). It is also easy to check (but
a nontrivial consistency check!) that the same is the case for our
$P$-branes, for which we have (see (\ref{vacstr}))
\begin{equation}
B^P_{m,n} = \left\{
\begin{array}{ll} 
\,\;\; 1 \qquad & \hbox{if $m$ and $n$ are even} \\
- \frac{1}{3} \quad & \hbox{otherwise.} 
\end{array}
\right.
\end{equation}
The fact that many of these coefficients $B_{m,n}$ are equal to $1$
suggests that the boundary state preserves in fact a large
W-algebra. This W-algebra is generated by all the $N=2$ primary
fields from the tensor product of the two vacuum representations
corresponding to $(m,n)$ with both $m$ and $n$ even. 

It is clear from the above arguments that these states define a
consistent W-algebra that is a proper subalgebra of the tensor
product of the two $N=2$ algebras. It is thus natural to ask whether
this property may be true for all generalised permutation branes. In
order to understand how this generalisation could work it is first
instructive to analyse the usual permutation branes.

\subsection{The W-algebra of the usual permutation branes}

If the levels of the two $N=2$ algebras are equal, the usual 
permutation branes are characterised by the property that
they preserve a large W-algebra ${\cal W}_{\rm per}$. This
W-algebra is the subalgebra of the  $(N=2)\times (N=2)$ 
algebra that consists of the states that are invariant under the
permutation of the two $N=2$ factors. Obviously, ${\cal W}_{\rm per}$
contains the diagonal $N=2$ algebra as a subalgebra, and we therefore
have  
\begin{equation}\label{Wein}
(N=2)_{\rm diag} \subset {\cal W}_{\rm per} \subset
(N=2) \times (N=2) \ . 
\end{equation}

In order to understand the structure of ${\cal W}_{\rm per}$ let us
decompose the tensor product of the two $N=2$ vacuum
representations with respect to the diagonal $N=2$ algebra  
\begin{equation} \label{tendecomp}
{\cal H}_0^{(1)}\otimes {\cal H}_0^{(2)} =
{\cal H}_{0,0} \oplus {\cal H}_{1,0} \oplus
{\cal H}_{2,0} \oplus 2 {\cal H}_{3,0} \oplus
{\cal H}_{7/2,1} \oplus {\cal H}_{7/2,-1} \oplus
3 {\cal H}_{4,0} \oplus \cdots \ .
\end{equation}
Here ${\cal H}_{H,Q}$ denotes the irreducible representation of the
diagonal $N=2$ algebra with conformal weight $H$ and charge $Q$, and
we have assumed that $k_1$ and $k_2$ are generic. For example, the
highest weight state of the diagonal $N=2$ algebra of the
representation ${\cal H}_{1,0}$ is  
\begin{equation}\label{genlev1}
K^{(1)}\equiv |1,0\rangle = 
\left( c_2 J^{(1)}_{-1} - c_1 J^{(2)}_{-1} \right) \Omega \ ,
\end{equation}
while the highest weight vector of ${\cal H}_{2,0}$ is 
\begin{align}
K^{(2)}\equiv |2,0\rangle = & \frac{1}{3}
\Big( c_2(1- c_2 ) 
\big( 2L^{(1)}_{-2} -3  (J^{(1)}_{-1})^{2}\big)\nonumber \\
& \quad + c_1(1- c_1 ) \big( 2L^{(2)}_{-2} -3 
(J^{(2)}_{-1})^{2} \big) \nonumber\\
& \quad - 6 (1- c_1 ) (1- c_2 ) 
 J^{(1)}_{-1} J^{(2)}_{-1}
  \Big) \Omega  \ , \label{genlev2}
\end{align}
and $\Omega$ denotes the vacuum vector.
The W-algebra ${\cal W}_{\rm per}$ now consists of those diagonal
$N=2$ representations for which the highest weight state is symmetric
under the exchange of the two $N=2$ algebras. Clearly 
$K^{(1)}$ is anti-symmetric (if $c_1=c_2$), while 
$K^{(2)}$ in (\ref{genlev2}) is symmetric (if $c_1=c_2$). 
Continuing in this fashion one finds that for generic $k_1=k_2$ the
W-algebra ${\cal W}_{\rm per}$ is  
\begin{equation}\label{Wperm}
{\cal W}_{\rm per} = {\cal H}_{0,0} \oplus 
{\cal H}_{2,0} \oplus {\cal H}_{3,0} \oplus 
{\cal H}_{7/2,1} \oplus {\cal H}_{7/2,-1} \oplus 2 {\cal H}_{4,0} 
\oplus \cdots \ . 
\end{equation}

\subsection{A W-algebra for the generalised permutation branes?} 

As is clear from the above, the decomposition (\ref{tendecomp}) also
holds if $k_1\ne k_2$. Given that the matrix factorisation description
of the generalised permutation branes is very similar to that of the
usual permutation branes, one may suspect that also the
generalised permutation branes preserve a large W-algebra
${\cal W}_{\rm gper}$. This W-algebra must again contain the diagonal
$N=2$ algebra, and we can therefore, as before, decompose it in terms
of representations of the diagonal $N=2$ algebra. Since 
${\cal W}_{\rm  gper}$ must be a subalgebra of the tensor product of
the two $N=2$ algebras, only the representations in (\ref{tendecomp})
may appear. 

Unlike the situation where the levels are equal, we do not know 
{\it a priori} how to characterise ${\cal W}_{\rm  gper}$. However, we
may analyse step by step whether a given $N=2$ representation may be
part of ${\cal W}_{\rm  gper}$ or not. For example, it is easy to see
that  ${\cal H}_{1,0}$ cannot appear in ${\cal W}_{\rm gper}$, since
$K^{(1)}$, together with the diagonal $N=2$ algebra, generates the
full tensor product algebra (upon taking
(anti-)commutators).\footnote{This is also the reason why the 
representation ${\cal H}_{1,0}$ cannot appear in the W-algebra
${\cal W}_{\rm per}$, see (\ref{Wperm}) above.} On the other hand, one
may at first think that $K^{(2)}$ could be part of 
${\cal W}_{\rm gper}$. However, as we shall now explain, this cannot
be the case. 

Using standard conformal field theory techniques we can express the
modes of $K^{(2)}_n$ in terms of the two $N=2$ algebras
\begin{align}
K^{(2)}_{n} = & \frac{1}{3}
\Big( c_2(1- c_2 ) 
\big( 2 L^{(1)}_{n} -3 \sum_{m} : J^{(1)}_{m+n}J^{(1)}_{-m}:\big)\nonumber \\
& \quad + c_1(1- c_1 ) \big( 2 L^{(2)}_{n} -3 
\sum_{m} : J^{(2)}_{m+n} J^{(2)}_{-m}: \big) \nonumber\\
& \quad - 6 (1- c_1 ) (1- c_2 ) 
\sum_{m} : J^{(1)}_{m+n} J^{(2)}_{-m}: \Big)
\ .
\end{align}
If $K^{(2)}$ is part of ${\cal W}_{\rm gper}$, then so is any state
that can be obtained by the action of $K^{(2)}$ modes and diagonal
$N=2$ modes from the vacuum. For example, we find that
\begin{equation}\label{prob}
G^+_{1/2}\, G^-_{1/2} \, K^{(2)}_0 \, K^{(2)} = 
a\,  J_{-1} \Omega + b\, K^{(1)} \ , 
\end{equation}
where
\begin{eqnarray}
a & = & \frac{8}{3 (c_1+c_2)}
(c_1+c_2-1)(c_1+c_2-2) 
(c_1-1) (c_2-1) c_1\, c_2 \ ,
\nonumber \\ 
b & = & -\frac{8}{c_1+c_2} 
(c_1-c_2) (c_1-1) (c_2-1) 
(c_1+c_2-1)^2  \ .
\end{eqnarray}
Unless $b=0$ it thus follows that (\ref{prob}) contains $K^{(1)}$, and
hence ${\cal W}_{\rm  gper}$ would in fact again be the full tensor
product. On the other hand, it is clear that the generalised permutation
branes cannot preserve the full tensor product symmetry. Thus we
conclude that $K^{(2)}$ cannot be part of ${\cal W}_{\rm  gper}$
unless $b=0$

On the other hand, $b$ only vanishes if either 
$c_1=c_2$ ($k_1=k_2$) or $c_{i}=1$
($k_i=1$).\footnote{Recall that $c=1$ is the smallest
possible value for the $N=2$ minimal models.}
The former case is just the usual permutation case for
which we have seen that $K^{(2)}$ is in fact part of 
${\cal W}_{\rm per}$. On the other hand, if $k_i=1$ the vector 
$K^{(2)}$ is simply a null-vector, and hence there is nothing to
discuss. (Incidentally this is also what happens in the analysis of
the $(k=1)\times (k=4)$ example.)

It is obviously conceivable that there exists an $N=2$ primary field
at higher conformal weight that can be consistently added to the
diagonal $N=2$ algebra to produce a proper subalgebra 
${\cal W}_{\rm  gper}$ of the tensor product of the two $N=2$
algebras. However, given the above result for $K^{(2)}$ this seems  
unlikely to us. We therefore suspect that the generalised permutation
branes are {\em not} characterised by the property that they preserve
{\em any extended $N=2$ symmetry}. 

It would be very interesting to understand how the generalised
permutation branes can be characterised in general. Given their 
simplicity from the matrix factorisation point of view they should be
singled out by some special property, but as we have just seen, this
does not seem to involve the symmetry they preserve. This suggests
that there must be a different point of view from which this class of
boundary states is preferred. A similar phenomenon also occurred for
the usual permutation branes \cite{Brunner:2005fv}, where it was found
that only a subset of the permutation factorisations corresponds in
fact to permutation branes. (The other permutation factorisations 
thus correspond to branes that preserve less symmetry.) The different
permutation factorisations (that have a uniform matrix factorisation
description) therefore correspond to branes that cannot be uniformly
characterised in terms of the symmetry they preserve.

\section{The $\boldsymbol{\hat{P}}$-branes}

For the $k_1=1$, $k_2=4$ theory there is a second natural class of
branes that define `generalised permutation branes', but that do not
correspond to the above matrix factorisations. 
These branes (which we shall denote as $\hat{P}$-branes in the
following) have in fact a simple description in terms of the
permutation orbifold of the $(k=1)^3$ theory that is also equivalent
to the $(k=1)\times (k=4)$ theory; for completeness we shall 
also briefly describe them from the various points of view.

\subsection{The conformal field theory description}

The $\hat{P}$-branes couple only to the sectors 
\begin{equation}
(\tfrac{1}{2}\tfrac{1}{2},\tfrac{1}{2}\tfrac{1}{2})\ , \
(\tfrac{1}{2}\tfrac{1}{2},\tfrac{5}{2}\tfrac{5}{2})\ , \
(\tfrac{3}{2}\tfrac{1}{2},\tfrac{1}{2}\tfrac{5}{2})\ , \
(\tfrac{1}{2}\tfrac{3}{2},\tfrac{5}{2}\tfrac{1}{2}) \ .
\end{equation}
In the first two sectors there is a non-trivial overlap with the
tensor product branes. The relevant combinations of Ishibashi states
with respect to the diagonal $N=2$ symmetry are then
\begin{align}
|\tfrac{1}{2}\tfrac{1}{2},\tfrac{1}{2}\tfrac{1}{2}\rrangle^{\hat{P}}\
=\ & |0,0\rrangle  + \sum_{m=1}^{\infty}\sum_{n=1}^{m} \cos \tfrac{2\pi
(m+n)}{3} |m^{2}+n^{2}-mn,0\rrangle  \nonumber \\
& \qquad + \sum_{n=0}^{\infty} \big(
|6n+\tfrac{11}{2},1\rrangle + |6n+\tfrac{11}{2},-1\rrangle  \big)
\label{phatvac}\\[1mm]
|\tfrac{1}{2}\tfrac{1}{2},\tfrac{5}{2}\tfrac{5}{2}\rrangle^{\hat{P}}\
=\ & \sum_{m=1}^{\infty}\sum_{n=1}^{m} \cos \tfrac{2\pi
(m+n)}{3} |m^{2}+n^{2}-mn,0\rrangle  \nonumber \\
& \qquad + \sum_{n=0}^{\infty} \big(
|6n+\tfrac{5}{2},1\rrangle + |6n+\tfrac{5}{2},-1\rrangle  \big) \ .
\end{align}
Their coefficients $B_{m,n}$ are thus of the form 
\begin{equation}
B^{\hat{P}}_{m,n} = \left\{
\begin{array}{ll} 
\,\;\; 1 \qquad & \hbox{if $m+n=0$ mod $3$} \\
- \frac{1}{2} \quad & \hbox{otherwise.} 
\end{array}
\right.
\end{equation}
These coefficients satisfy again the factorisation 
constraint (\ref{facto}). Since now $B_{m,n}=1$ for $m+n=0$ mod $3$,
these branes preserve a W-algebra that is different from the one for
the $P$-branes. 

In the other two sectors the branes overlap with the generalised
permutation branes. The relevant combinations of Ishibashi states for
the $\hat{P}$-branes are
\begin{align}
|\tfrac{3}{2}\tfrac{1}{2},\tfrac{1}{2}\tfrac{5}{2}\rrangle^{\hat{P}}
= & \sum_{m=0}^{\infty}\sum_{\substack{n=0\\
\makebox[0cm][c]{$\scriptstyle\!\!\!\!\!\text{$m+1$ and $n$
    even}$}}}^{\infty}| \tfrac{1}{3}+ (m-n)^{2}+ (m+1)n,0\rrangle \nonumber\\
& + \sum_{m=0}^{\infty}\sum_{\substack{n=0\\
\makebox[0cm][l]{$\mspace{-50mu}\scriptstyle  \text{$m+1$ or $n$ odd}$}\\ 
\makebox[0cm][l]{$\mspace{-50mu}\scriptstyle m+n+\text{Mod} (m,2)
+\text{Mod} (n,2)=0 \mod 3 $}}}^{\infty} 
|\tfrac{1}{3}+ (m-n)^{2}+ (m+1)n,0\rrangle \nonumber\\
& + \sum_{m=0}^{\infty}\sum_{\substack{n=0\\
\makebox[0cm][l]{$\mspace{-50mu}\scriptstyle  \text{$m+1$ or $n$ odd}$}\\ 
\makebox[0cm][l]{$\mspace{-50mu}\scriptstyle m+n+\text{Mod} (m,2)
+\text{Mod} (n,2)=1 \mod 3 $}}}^{\infty} 
\Big(-\frac{\sqrt{3}+1}{2} \Big)
|\tfrac{1}{3}+ (m-n)^{2}+ (m+1)n,0\rrangle \nonumber\\
& + \sum_{m=0}^{\infty}\sum_{\substack{n=0\\
\makebox[0cm][l]{$\mspace{-50mu}\scriptstyle  \text{$m+1$ or $n$ odd}$}\\ 
\makebox[0cm][l]{$\mspace{-50mu}\scriptstyle m+n+\text{Mod} (m,2)
+\text{Mod} (n,2)=2 \mod 3 $}}}^{\infty} 
\Big(\frac{\sqrt{3}-1}{2} \Big)
|\tfrac{1}{3}+ (m-n)^{2}+ (m+1)n,0\rrangle \ ,
\end{align}
and similarly for
$|\tfrac{1}{2}\tfrac{3}{2},\tfrac{5}{2}\tfrac{1}{2}\rrangle^{\hat{P}}$.
Here, $\text{Mod} (m,2)$ is $0$ or $1$ depending on whether $m$ is
even or odd, respectively. The corresponding boundary states read 
\begin{equation}
|\!|\hat{P} (j)\rrangle \ =\ \sqrt{2}\cdot 3^{\frac{1}{4}}
\Big(|\tfrac{1}{2}\tfrac{1}{2},\tfrac{1}{2}\tfrac{1}{2}\rrangle^{\hat{P}}
+ |\tfrac{1}{2}\tfrac{1}{2},\tfrac{5}{2}\tfrac{5}{2}\rrangle^{\hat{P}}\Big)
+  3^{\frac{1}{4}} e^{2\pi i j/3}
|\tfrac{3}{2}\tfrac{1}{2},\tfrac{1}{2}\tfrac{5}{2}\rrangle^{\hat{P}} +
  3^{\frac{1}{4}} e^{-2\pi i j/3}
|\tfrac{1}{2}\tfrac{3}{2},\tfrac{5}{2}\tfrac{1}{2}\rrangle^{\hat{P}}
 \ ,
\end{equation}
where $j=0,1,2$.

\noindent The self-overlap of these boundary states is given by 
\begin{align}
\llangle \hat{P} (j)|\!|
& q^{\frac{1}{2} (L_{0}+\tilde{L}_{0})-\frac{c}{24}} |\!|
\hat{P} (j)\rrangle \nonumber\\
=  & \mmch[\tilde{q}]{1}{1}{1} \Big(\mmch[\tilde{q}]{4}{1}{1} +
\mmch[\tilde{q}]{4}{5}{5} + 
2\mmch[\tilde{q}]{4}{5}{1} + 2\mmch[\tilde{q}]{4}{1}{5} \Big) \nonumber \\
& + \mmch[\tilde{q}]{1}{1}{3}
\Big(\mmch[\tilde{q}]{4}{1}{9}+\mmch[\tilde{q}]{4}{7}{3} + 2
\mmch[\tilde{q}]{4}{3}{3} 
+2\mmch[\tilde{q}]{4}{1}{5} \Big) \nonumber \\
& + \mmch[\tilde{q}]{1}{3}{1}
\Big(\mmch[\tilde{q}]{4}{9}{1}+\mmch[\tilde{q}]{4}{3}{7} + 2
\mmch[\tilde{q}]{4}{3}{3} 
+2\mmch[\tilde{q}]{4}{5}{1} \Big) \nonumber \\
& + 
\sum_{m=1}^{\infty}\sum_{\substack{n=1\\
    \makebox[0cm][c]{$\scriptstyle\!\!\!\!\!m+n \not\equiv 0
      \,\text{mod}\, 3$}}}^{m} 
\ \dgch[\tilde{q}]{\frac{1}{9} (m^{2}+n^{2}-mn)}{0} \nonumber\\
=\ & \big(\mmch[\tilde{q}]{1}{1}{1}\mmch[\tilde{q}]{1}{1}{1} +
\mmch[\tilde{q}]{1}{3}{1}\mmch[\tilde{q}]{1}{3}{1} +
\mmch[\tilde{q}]{1}{1}{3}\mmch[\tilde{q}]{1}{1}{3}   \big)\nonumber\\
& \times \big(\mmch[\tilde{q}]{1}{1}{1}+\mmch[\tilde{q}]{1}{3}{1}
+\mmch[\tilde{q}]{1}{1}{3} \big) \nonumber\\
& + \big( 
\mmch[\tilde{q}^{\frac{1}{3}}]{1}{1}{1}
+\mmch[\tilde{q}^{\frac{1}{3}}]{1}{1}{3}
+\mmch[\tilde{q}^{\frac{1}{3}}]{1}{3}{1} \big) \ .
\label{nn1}
\end{align}
In the last step we have used the identities of appendix~B.3 to
rewrite the spectrum in terms of $k=1$-characters.

The $\hat{P}$-boundary states are in fact elementary although we
are only considering the NSNS sector. This implies that they actually
do not couple to {\it any} RR states. In particular, they therefore
cannot carry any RR charges.

\subsection{The matrix factorisation point of view}

Since the $\hat{P}$-branes do not couple to the RR sector, the
$\hat{P}$-branes coincide with their own anti-branes. In the 
matrix factorisation language this means that the corresponding
factorisations must be equivalent to their reverse
factorisations. We now propose that the relevant factorisations are 
\begin{equation}
E = \left( \begin{matrix}
(x-\xi y^{2}) & y^3 \\ -y^3 & \pi_\xi(x,y^{2}) 
\end{matrix} \right) \ , 
\qquad
J = \left( \begin{matrix}
\pi_\xi(x,y^{2}) & -y^3 \\ y^3 & (x-\xi y^{2})
\end{matrix} \right) \ , 
\label{phatfac}
\end{equation}
where we have used the notation 
\begin{equation}\label{defofpi}
\pi_{\xi} (v,w) = \prod_{\xi'\not= \xi} (v-\xi' w) \ ,
\end{equation}
and $\xi$ is again a third root of $-1$ (see eq.\
\eqref{genpermfactexample}). It is very easy to see that this
factorisation is indeed equivalent to its own reverse.

It follows from (\ref{nn1}) that the self-spectrum of the
$\hat{P}$-branes contains six bosonic (and six  
fermionic) states, whose $U(1)$ charges are 
$0,\tfrac{1}{3},\tfrac{1}{3},\tfrac{2}{3},\tfrac{2}{3},1$. 
This can now be compared with the self-spectrum of the corresponding
matrix factorisation (\ref{phatfac}). Because the factorisation is
equivalent to its own reverse, it is sufficient to consider only the
bosons. The boson with $U(1)$ charge $0$ is simply the identity
matrix, $\phi_0=\phi_1={\bf 1}$, and the boson with 
$U(1)$ charge $1$ can be represented by the off-diagonal matrix
\begin{equation}
\phi_1 = \left( \begin{matrix}
0 & (x -2 y^2 ) y^2 \\ \quad -3x - 4 y^2  \quad & 0 
\end{matrix} \right) \ , 
\end{equation}
where, for simplicity, we have only given the result for
$\xi=-1$. [The matrix $\phi_0$ is then determined from $\phi_1$ as 
$\phi_0 = \frac{1}{W} E \phi_1 J$.] For $\xi=-1$, the two bosons of
charge $\tfrac{1}{3}$ can be described, up to exact solutions, by the
matrix 
\begin{equation}
\phi_1 = \left( \begin{matrix}
ay & b \, (-x+2 y^2 ) \\
\quad b \quad & (a -3b) y
\end{matrix} \right) \ , 
\end{equation}
while the two bosons of charge $\tfrac{2}{3}$ can (again for
$\xi=-1$) be described by the matrix
\begin{equation}
\phi_1 = \left( \begin{matrix}
ax & bxy \\\quad -by \quad & ax + b(-2x+y^2) 
\end{matrix} \right) \ .
\end{equation}
The result for the other values of $\xi$ is similar. This gives
strong support to the claim that (\ref{phatfac}) is indeed the matrix 
factorisation of the $\hat{P}$-branes.

\subsection{The permutation orbifold point of view}

The $\hat{P}$-branes have in fact a simple description in terms of
yet another realisation of the $A_1\times A_4$ model, namely as a
permutation orbifold of three minimal models at $k=1$. The 
relevant triple product of $k=1$ theories is the one that
corresponds to the superpotential 
\begin{equation}\label{WA1A1A1}
W=W_{A_{1} \times A_{1} \times A_{1}'} = u^{3}+v^{3}+w^{3}+z^{2} \ .
\end{equation}
Here $A_{1}'$ denotes the $A_1$ minimal model with the opposite GSO
projection \cite{Kapustin:2003rc} (see also \cite{Brunner:2005pq}).
Obviously, all the different branes we have discussed in this paper
can also be described in terms of matrix factorisations of $W$ in 
(\ref{WA1A1A1}); this is discussed in appendix~C. 

To obtain the $A_{1}\times A_{4}$ model from this theory, we have to
perform a $\mathbb{Z}_{2}$-orbifold which acts as $v\leftrightarrow w$   
together with $z\to -z$. Since the $u^3$ term is not involved,
this just means that the $\mathbb{Z}_2$ orbifold of $A_1\times A_1'$
is in fact equivalent to the $A_4$ model. This can, for example, be
confirmed by checking that the relevant matrix factorisations are in
one-to-one correspondence.
\smallskip

\noindent The conformal field theory corresponding to (\ref{WA1A1A1})
is 
\begin{equation}
\mathcal{H} = \bigoplus_{[l_{i},m_{i},s_{i}],s'_{i}} \bigotimes_{i=1}^{3} 
\mathcal{H}_{[l_{i},m_{i},s_{i}]} \otimes
\mathcal{H}_{[l_{i},m_{i},s'_{i}]}  \ ,
\end{equation}
where $i=1,2,3$ labels the sectors corresponding to the variables
$u,v,w$. The sum over the $s_{i}$ and $s'_{i}$ is restricted such
that they are either all odd or all even and by the requirement that
\begin{equation}\label{GSOA13}
s_{1}+s_{2}+s_{3}-s'_{1}-s'_{2}-s'_{3} = 0 \mod 4 \ .
\end{equation}
We note in passing that this GSO projection is incompatible with the
B-type gluing condition in the Ramond-Ramond sector. To see this we
recall that the B-type condition requires the left- and right-moving
$U (1)$-charge to be related as $Q_{L}=-Q_{R}$. Expressing
the charges in terms of the coset labels (see appendix~A),  it follows that 
\begin{equation}
\frac{2}{3} (m_{1}+m_{2}+m_{3}) - \frac{1}{2}
(s_{1}+s_{2}+s_{3}+s'_{1}+s'_{2}+s'_{3}) = 0 \mod 2 \ .
\end{equation}
Using the constraint~\eqref{GSOA13} B-type gluing would require
\begin{equation}
\frac{2}{3} (m_{1}+m_{2}+m_{3}) = 1 \mod 2 \ .
\end{equation}
For integer labels $m_{i}$ this can never be achieved, and thus all
B-type branes of this theory only couple to the NSNS sector.

The analogue of the above $\mathbb{Z}_2$-orbifold in the
Landau-Ginzburg description is now the $\mathbb{Z}_{2}$-orbifold that
acts as a transposition in the second and third factor. 
From the character decompositions in appendix B.3 this is fairly
obvious. It is also not hard to see how the GSO-projections are
related: the projection~\eqref{GSOA13} allows {\it e.g.}\ the labels
$s_{i}$ to equal $s'_{i}$. The decomposition into $k=4$
representations  and restriction to the invariant part leaves us again
with identical left- and right-moving labels $s$. On the $(k=1)\times
(k=4)$ side we thus obtain the GSO condition
\begin{equation}
s_{1}+S_{2}-s'_{1}-S'_{2} =0 \mod 4 \ ,
\end{equation}
where we denoted the coset label of the $(k=4)$-part by (capital)
$S_{2}$. This is the same GSO projection as in~\eqref{k4k1spec}.
\medskip

From the permutation orbifold point of view the $\hat{P}_{j}$-branes
are simply the superposition of  
the two permutation boundary states $|\!|(12)_{\xi_{j}}\rrangle$ and
$|\!|(13)_{\xi_{j}}\rrangle$. The self-spectrum of the
$\hat{P}$-branes thus consists of the self-spectrum of a transposition  
brane, 
\begin{align}
\llangle (12)_{\xi_{j}} |\!| q^{\frac{1}{2}
(L_{0}+\tilde{L}_{0})-\frac{c}{24}} |\!| (12)_{\xi_{j}}\rrangle
= & \big(\mmch[\tilde{q}]{1}{1}{1}\mmch[\tilde{q}]{1}{1}{1} +
\mmch[\tilde{q}]{1}{3}{1}\mmch[\tilde{q}]{1}{3}{1} \nonumber \\
& + \mmch[\tilde{q}]{1}{1}{3}\mmch[\tilde{q}]{1}{1}{3}   \big)
\nonumber \\
& \times \big(\mmch[\tilde{q}]{1}{1}{1}+\mmch[\tilde{q}]{1}{3}{1}
+\mmch[\tilde{q}]{1}{1}{3} \big)
\label{sspectrumtranspbrane}
\end{align}
and the relative spectrum between two transposition branes,
\begin{equation}\label{rspectrumtranspbranes}
\llangle (12)_{\xi_{j}} |\!| q^{\frac{1}{2}
(L_{0}+\tilde{L}_{0})-\frac{c}{24}} |\!| (13)_{\xi_{j}}\rrangle
=  \mmch[\tilde{q}^{\frac{1}{3}}]{1}{1}{1}
+\mmch[\tilde{q}^{\frac{1}{3}}]{1}{1}{3}
+ \mmch[\tilde{q}^{\frac{1}{3}}]{1}{3}{1}\ .
\end{equation}
The $\tilde{q}^{\frac{1}{3}}$ powers appear since the relative
permutation is of order three. This now agrees precisely with 
(\ref{nn1}). 

We also know that the $T(0,0)$ brane comes from the permutation brane 
$|\!|(23)_{\xi_0}\rrangle$ in the $(k=1)^3$ theory. It is invariant
under the $\mathbb{Z}_2$ orbifold, and decomposes into two fractional
branes corresponding to the $T(0,0)$ brane and its anti-brane. The
spectrum between the $T(0,0)$ brane and a $\hat{P}$ brane is then 
just given by (\ref{rspectrumtranspbranes}). Finally, it is also not
difficult to construct these branes on the torus --- they correspond to
superpositions of D1-branes at angles $\phi_n$. In contradistinction
to the $P$-branes, these D1-branes now do not go through any fixed
points and therefore need not be resolved.

\section{Conclusions}

In this paper we have constructed the generalised $N=2$ permutation
branes for the simple example of the product theory with levels $k=1$
and $k=4$. This was done by analysing the product theory with respect
to the diagonal $N=2$ algebra. In particular, we classified all the 
B-type Ishibashi states with respect to the diagonal $N=2$ algebra,
and then constructed the generalised permutation boundary states by
taking suitable combinations of these Ishibashi states (see 
(\ref{final})). We then checked that these $P$-brane boundary states
reproduce the correct (topological) open string spectrum. We also
showed that they correspond to certain standard D-branes in the 
$\mathbb{Z}_6$-orbifold of the $su(3)$ torus, thus giving strong
support for their consistency. Finally, we also analysed a second
class of $\hat{P}$-brane boundary states, and explained how
they can be naturally described from the point of view of the
permutation orbifold of the product theory of three $k=1$
models. 

As we mentioned in the introduction, these D-branes play an important
role in accounting for the RR charges of Gepner models. Indeed, the
above constructions appear for $3$ of the $31$ A-type Gepner models
for which generalised permutation branes are required: these are the 
models $\mathbb{P}_{\left(3,3,4,6,8\right)}[24]$,
$\mathbb{P}_{\left(3,5,6,6,10\right)}[30]$ and 
$\mathbb{P}_{\left(3,10,12,15,20\right)}[60]$ (see table~2 of 
\cite{Caviezel:2005th}). For these Gepner models our construction
therefore gives a conformal field theory description of the missing
D-branes. 

It would obviously be very interesting to generalise the construction
to the other cases of interest. Clearly, the case of $c=3$ is special
since we have a free field realisation (as well as the relation to the
triple product theory), but the first approach (section~2) should also
be possible in the general case. In particular, the spectral flow
symmetry should still be useful in constraining the coefficients in
front of the different Ishibashi states. On the other hand, as we have
argued in section~3.7, it seems unlikely that the generalised
permutation branes preserve any extended $N=2$ symmetry, and it may
therefore be difficult to guess the correct ansatz.

It is clear that the generalised permutation branes are very special
from the point of view of the matrix factorisation
analysis. It would therefore be very interesting to understand how to
characterise these branes intrinsically in conformal field theory.
This will probably require a new way of looking at D-branes. One
natural guess would be that these branes are singled out by special
properties of their open string spectra. As we have seen in the
$(k=1)\times (k=4)$ case, the open string spectra look often much
simpler than the formulae for the boundary states. For example,
the spectrum of a single $P$-brane can still be written in terms of
characters of the full product algebra, although the boundary states
of the $P$-branes explicitly break the full $(N=2)\times (N=2)$
symmetry. It would be very interesting to understand which properties of 
the open string spectra will generalise.
\smallskip

The $N=2$ minimal models can be described as cosets of $SU(2)$
theories, and the generalised permutation branes of the $N=2$ theories
should therefore be closely related to the generalised permutation
branes for products of $SU(2)$ group
manifolds~\cite{Fredenhagen:2005an}. In the group case their existence
was argued for on the level of the effective world-volume
theory.\footnote{A first step to perform a similar analysis for coset
theories was also recently taken in~\cite{Sarkissian:2006xp}.} Given
the close relation between the $N=2$ algebra and $SU(2)$ one might
then expect that one can construct the generalised permutation branes
for the product of two $SU(2)$ WZW models at $k=1$ and $k=4$ by
combining our $N=2$ construction with suitable boundary conditions for
the $U(1)$ factors. The resulting branes on the two $SU(2)$s factorise
into the coset and a $U(1)$ part, and one can show that they do not
preserve the diagonal $SU(2)$ symmetry. This symmetry, however, is
expected from the analysis of~\cite{Fredenhagen:2005an}, and
it thus seems that we do not obtain the 
generalised permutation branes of~\cite{Fredenhagen:2005an}
in this manner. Nevertheless, we expect that the two constructions are
in fact closely related; it would be very interesting to understand
this better.

\section*{Acknowledgements}

This research has been partially supported by 
the Swiss National Science Foundation and the Marie Curie network
`Constituents, Fundamental Forces and Symmetries of the Universe'
(MRTN-CT-2004-005104). We thank Ilka Brunner and Thomas Quella for
useful discussions.

\section*{Appendix} 

\appendix

\section{Conventions}

A $N=2$ minimal model of level $k$ has central charge
\[
c=\frac{3k}{k+2}\ .
\]
These models can be described by a coset construction. The sectors of
the coset theory are labelled by three integers $(l,m,s)$ where
$l=0,\dotsc ,k$ and $m$ and $s$ are defined modulo $2 (k+2)$ and $4$,
respectively. Here $l+m+s$ is even, and we have the field
identifications $(l,m,s)\sim(k-l,m+k+2,s+2)$. These sectors are
representations of the bosonic subalgebra of the $N=2$ superconformal
algebra. 

In the following we will often use a different notation to label
the full representations of the $N=2$ algebra.
The Neveu-Schwarz sectors are labelled by two half-integers $u,v$,
where $0\leq u,v,u+v\leq k+1$. Conformal weight $h$ and $U (1)$ charge
$q$ are given by
\begin{equation}\label{confweightcharge}
h^{\text{NS}}_{(u,v)}=\frac{uv-1/4}{k+2}\quad \quad 
\quad q^{\text{NS}}_{(u,v)}=\frac{u-v}{k+2} \ .
\end{equation}
The relation to the coset sectors is
\begin{equation}
(u,v) \equiv (l,m,0) \oplus (l,m,2) \ , 
\end{equation}
where
\begin{equation}
l=u+v-1 \quad \quad \quad m=v-u\ .
\end{equation}

The Ramond sectors are labelled by two integers
$u,v$ with the range $0\leq v\leq k$ and $1\leq u\leq k+1-v$. The
conformal weight and charge is
\begin{align}
h^{\text{R}}_{(u,v)} &= \frac{uv}{k+2}+\frac{c}{24} & 
q^{\text{R}}_{(u,v)} &= \frac{u-v}{k+2}-\frac{1}{2} \ .
\end{align}
The relation to the coset sectors is
\begin{equation}
(u,v)\ \equiv \ (l,m,-1) \oplus (l,m,1) \ , 
\end{equation}
where again
\begin{equation}
l=u+v-1 \quad \quad \quad m=v-u\ .
\end{equation}
The representations with $v=0$ have Ramond ground states of charge
$q^{\text{R}}_{(u,v)}$. For the other representations, the charges of
the lowest lying states are $q^{\text{R}}_{(u,v)}$ and
$q^{\text{R}}_{(u,v)}+1$.

The $N=2$ algebra possesses a family of outer automorphisms, usually
referred to as spectral flow, that are parametrised by $t$. They act
on the generators of the algebra as 
\begin{eqnarray}
\alpha_t(L_n) & = & L_n - t \, J_n +\tfrac{t^2 c}{6}\, \delta_{n,0}
\nonumber 
\\
\alpha_t(J_n) & = & J_n - \tfrac{t c}{3}\, \delta_{n,0} 
\label{spectralflowofalgebra}
\\
\alpha_t(G^\pm_r) & = & G^\pm_{r\mp t} \ .\nonumber 
\end{eqnarray}
If $t$ is half-integer, the spectral flow connects 
Neveu-Schwarz- and Ramond-representations, whereas integer $t$ 
maps Neveu-Schwarz (Ramond) representations to themselves. It follows
immediately from the above that the conformal weight and charge of any
state transforms as 
\begin{equation}\label{spectralflow}
(h,q)\rightarrow (h-t\, q+\tfrac{t^2 c}{6},q-\tfrac{t c}{3}) \ .
\end{equation}
In general, a highest weight state of the original $N=2$ algebra is
however not highest weight with respect to the spectrally flowed
algebra; thus one cannot directly read off from
\eqref{spectralflow} the weight and charge of the highest
weight state of the new (spectrally flowed) representation. In the
generic case (when there is no `accidental null vector'), the
representation $(h_{t },q_{t })$ that is  reached by spectral
flow is  
\begin{equation}\label{genspectralflow}
(h_{t},q_{t}) \ =\ 
(h+t^{2} (\tfrac{c}{6}-\tfrac{1}{2})-t \, q,q-t
(\tfrac{c}{3}-1))\ . 
\end{equation}
For $c=3$ this simplifies to the formula (valid in the generic case) 
\begin{equation}\label{c=3spectralflow}
(h_{t},q_{t}) \ =\ 
(h-t \, q,q)\ . 
\end{equation}
For example, for $t=-3$, this then maps the representation 
$(6n+\tfrac{5}{2},1) \mapsto (6n+\tfrac{11}{2},1)$. On the other hand, 
(\ref{c=3spectralflow}) does not always hold; for example, for the
case of the vacuum representation, there is an `accidental' null
vector so that the original ground state is also highest weight with
respect to the $t=-1$ algebra. Thus under $t=-1$, 
we have $(0,0)\mapsto (\tfrac{1}{2},1)$, leading to the 
flow $(0,0)\mapsto (\tfrac{5}{2},1)$ for $t=-3$.

For the $k^{\rm th}$ minimal model, it is easy to see that spectral
flow by $t=\pm(k+2)$ acts trivially on all representations. Thus we
may restrict to $0\leq t < k+2$, for which we find
\begin{equation}
(u,v)\rightarrow \left\{\begin{array}{ll}
(u+t,v-t) & \qquad \text{ for } v\geq t \\
(t-v,k+2-u-t) & \qquad \text{ for }v<t \ .
\end{array}
\right.
\end{equation}

\section{Decomposition of characters}

\subsection{$k_{1}=1$, $k_{2}=4$, Neveu-Schwarz sector} 
In this appendix we list the decomposition of characters of the tensor
product of two minimal models with level $k_{1}=1$ (central charge
$c_{1}=1$) and with level $k_{2}=4$ ($c_{2}=2$) into characters of the
diagonal $N=2$ algebra. The labelling of the sectors is done by
half-integers $u_{1},v_{1},u_{2},v_{2}$; the conformal weights and
charges are determined as in~\eqref{confweightcharge}.

\noindent The decomposition of the characters is given by
\begin{align}
\mmch{1}{1}{1}\mmch{4}{1}{1}\ =\ & \dgch{0}{0}+ \sum_{m=1}^{\infty}
\sum_{n=1}^{m} \dgch{(m-n)^{2}+mn}{0}\nonumber \\
& + \sum_{n=0}^{\infty}\left( \dgch{6n+\frac{11}{2}}{1} +
\dgch{6n+\frac{11}{2}}{-1} \right) \label{vacdec1}\\   
\mmch{1}{1}{1}\mmch{4}{3}{3}\ =\ & \sum_{m=0}^{\infty}
\sum_{n=0}^{\infty} \dgch{\frac{1}{3}+ (m-n)^{2}+ (m+1)n}{0}\\
\mmch{1}{1}{1}\mmch{4}{5}{5}\ =\ & \sum_{m=1}^{\infty}
\sum_{n=1}^{m} \dgch{(m-n)^{2}+mn}{0}\nonumber \\
& + \sum_{n=0}^{\infty} \left( \dgch{\frac{5}{2}+6n}{1} +
\dgch{\frac{5}{2}+6n}{-1} \right) \\
\mmch{1}{3}{1}\mmch{4}{1}{5}\ =\ & \sum_{m=0}^{\infty}
\sum_{n=0}^{\infty} \dgch{\frac{1}{3}+ (m-n)^{2}+ (m+1)n}{0}\\
\mmch{1}{3}{1}\mmch{4}{3}{7}\ =\ & \sum_{m=1}^{\infty}
\sum_{n=1}^{m} \dgch{(m-n)^{2}+mn}{0}\nonumber \\
& + \sum_{n=0}^{\infty} \left( \dgch{\frac{3}{2}+6n}{-1} +
\dgch{\frac{7}{2}+6n}{1} \right) \\
\mmch{1}{3}{1}\mmch{4}{9}{1}\ =\ &\sum_{n=0}^{\infty}
\left( \dgch{\frac{1}{2}+6n}{1} + \dgch{\frac{9}{2}+6n}{-1} \right)\nonumber
\\
& + \sum_{m=1}^{\infty} \sum_{n=1}^{m} \dgch{(m-n)^{2}+mn}{0} \ .
\end{align}
These are the ones that are relevant for B-type boundary
conditions. The others are
\begin{align}
\mmch{1}{3}{1}\mmch{4}{1}{1}\ =\ & \sum_{n=0}^{\infty} 
\Big( \dgch{\frac{1}{6}+2n}{\frac{1}{3}} + 
\dgch{\frac{1}{6}+\frac{7}{2}+4n}{-\frac{2}{3}} \Big)\\
\mmch{1}{1}{1}\mmch{4}{3}{1}\ =\ & \sum_{n=0}^{\infty} 
\Big( \dgch{\frac{1}{12}+n}{\frac{1}{6}} +
\dgch{\frac{1}{12}+\frac{9}{2}+5n}{-\frac{5}{6}} \Big) \\
\mmch{1}{1}{1}\mmch{4}{5}{1}\ =\ & \sum_{n=0}^{\infty} 
\Big( \dgch{\frac{1}{6}+n}{\frac{1}{3}} +
\dgch{\frac{1}{6}+\frac{3}{2}+2n}{-\frac{2}{3}} \Big)
\end{align}
\begin{align}
\mmch{1}{1}{1}\mmch{4}{7}{1}\ =\ & \sum_{n=0}^{\infty} 
\Big( \dgch{\frac{1}{4}+3n}{\frac{1}{2}} +
\dgch{\frac{1}{4}+1+3n}{\frac{1}{2}} \nonumber \\
& \qquad +
\dgch{\frac{1}{4}+\frac{3}{2}+3n}{-\frac{1}{2}}  +
\dgch{\frac{1}{4}+\frac{5}{2}+3n}{-\frac{1}{2}} \Big) \\
\mmch{1}{1}{1}\mmch{4}{9}{1}\ =\ & \sum_{n=0}^{\infty} 
\Big( \dgch{\frac{1}{3}+4n}{\frac{2}{3}} +
\dgch{\frac{1}{3}+\frac{3}{2}+2n}{-\frac{1}{3}} \Big)\\
\mmch{1}{1}{1}\mmch{4}{5}{3}\ =\ & \sum_{n=0}^{\infty} 
\Big( \dgch{\frac{7}{12}+n}{\frac{1}{6}} +
\dgch{\frac{7}{12}+\frac{3}{2}+5n}{-\frac{5}{6}} \Big)\\
\mmch{1}{1}{1}\mmch{4}{7}{3}\ =\ & \sum_{n=0}^{\infty} 
\Big( \dgch{\frac{5}{6}+2n}{\frac{1}{3}} +
\dgch{\frac{5}{6}+\frac{3}{2}+4n}{-\frac{2}{3}} \Big)\\
\mmch{1}{3}{1}\mmch{4}{3}{3}\ =\ & \sum_{n=0}^{\infty} 
\Big( \dgch{\frac{1}{2}+n}{\frac{1}{3}} +
\dgch{\frac{1}{2}+\frac{1}{2}+2n}{-\frac{2}{3}} \Big)\\
\mmch{1}{3}{1}\mmch{4}{5}{5}\ =\ & \sum_{n=0}^{\infty} 
\Big( \dgch{\frac{7}{6}+2n}{\frac{1}{3}} +
\dgch{\frac{7}{6}+\frac{1}{2}+4n}{-\frac{2}{3}} \Big)\\
\mmch{1}{3}{1}\mmch{4}{3}{1}\ =\ & \sum_{n=0}^{\infty} 
\Big( \dgch{\frac{1}{4}+3n}{\frac{1}{2}} +
\dgch{\frac{1}{4}+2+3n}{\frac{1}{2}} \nonumber \\
& \qquad + 
\dgch{\frac{1}{4}+\frac{1}{2}+3n}{-\frac{1}{2}} + 
\dgch{\frac{1}{4}+\frac{5}{2}+3n}{-\frac{1}{2}}  \Big) \\
\mmch{1}{3}{1}\mmch{4}{1}{3}\ =\ & \sum_{n=0}^{\infty} 
\Big( \dgch{\frac{1}{4}+n}{\frac{1}{6}} +
\dgch{\frac{1}{4}+\frac{7}{2}+5n}{-\frac{5}{6}} \Big)\\
\mmch{1}{3}{1}\mmch{4}{5}{1}\ =\ & \sum_{n=0}^{\infty} 
\Big( \dgch{\frac{1}{3}+2n}{\frac{2}{3}} +
\dgch{\frac{1}{3}+\frac{1}{2}+n}{-\frac{1}{3}} \Big)\\
\mmch{1}{3}{1}\mmch{4}{7}{1}\ =\ & \sum_{n=0}^{\infty} 
\Big( \dgch{\frac{5}{12}+5n}{\frac{5}{6}} +
\dgch{\frac{5}{12}+\frac{1}{2}+n}{-\frac{1}{6}} \Big)\\
\mmch{1}{3}{1}\mmch{4}{1}{7}\ =\ & \sum_{n=0}^{\infty} 
\Big( \dgch{\frac{5}{12}+n}{-\frac{1}{6}} +
\dgch{\frac{5}{12}+\frac{5}{2}+5n}{\frac{5}{6}} \Big)\\
\mmch{1}{3}{1}\mmch{4}{5}{3}\ =\ & \sum_{n=0}^{\infty} 
\Big( \dgch{\frac{3}{4}+3n}{\frac{1}{2}} +
\dgch{\frac{3}{4}+1+3n}{\frac{1}{2}} \nonumber \\
& \qquad +  
\dgch{\frac{3}{4}+\frac{1}{2}+3n}{-\frac{1}{2}} +  
\dgch{\frac{3}{4}+\frac{3}{2}+3n}{-\frac{1}{2}} \Big) \\
\mmch{1}{3}{1}\mmch{4}{3}{5}\ =\ & \sum_{n=0}^{\infty} 
\Big( \dgch{\frac{3}{4}+n}{\frac{1}{6}} +
\dgch{\frac{3}{4}+\frac{1}{2}+5n}{-\frac{5}{6}} \Big) \\
\mmch{1}{3}{1}\mmch{4}{7}{3}\ =\ & \sum_{n=0}^{\infty} 
\Big( \dgch{1+4n}{\frac{2}{3}} +
\dgch{1+\frac{1}{2}+2n}{-\frac{1}{3}} \Big)\\
\mmch{1}{3}{1}\mmch{4}{1}{9}\ =\ & \sum_{n=0}^{\infty} 
\Big( \dgch{\frac{1}{2}+2n}{-\frac{1}{3}} +
\dgch{\frac{1}{2}+\frac{5}{2}+4n}{\frac{2}{3}} \Big) \ .
\end{align}
There are some identities between these products of characters which
will be useful later, namely
\begin{align}\label{identitiess}
\mmch{1}{3}{1} \Big( \mmch{4}{1}{1} + \mmch{4}{5}{5} \Big)\ 
=&\ \mmch{1}{1}{1}\mmch{4}{5}{1}\\
\mmch{1}{3}{1} \Big( \mmch{4}{1}{9} + \mmch{4}{7}{3} \Big)\ 
=&\ \mmch{1}{1}{3}\mmch{4}{3}{3}\\
\mmch{1}{1}{1} \Big( \mmch{4}{9}{1} + \mmch{4}{3}{7} \Big)\ 
=&\ \mmch{1}{3}{1}\mmch{4}{5}{1} 
\end{align}
\begin{equation}
\mmch{1}{1}{1}\mmch{4}{3}{3}\ =\ \mmch{1}{3}{1}\mmch{4}{1}{5}\ =\
\mmch{1}{1}{3}\mmch{4}{5}{1}\ . 
\end{equation}
Furthermore we have
\begin{align}
6 \sum_{n\geq 1}\sum_{m\geq n} \dgch{m^{2}+n^{2}-mn}{0} +
\frac{\vartheta_{3} (z,q)}{\eta^{3} (q)} = &
\mmch{1}{1}{1} \Big(\mmch{4}{1}{1}+\mmch{4}{5}{5}\Big)\nonumber\\
& + \mmch{1}{3}{1}
\Big(\mmch{4}{9}{1}+\mmch{4}{3}{7}\Big) \nonumber\\
&+ \mmch{1}{1}{3}
\Big( \mmch{4}{1}{9}+\mmch{4}{7}{3} \Big) \ .
\label{identitiese}
\end{align}

\subsection[$k_{1}=1,k_{2}=4$, Ramond sector]{$\boldsymbol{k_{1}=1,k_{2}=4}$, 
Ramond sector}
The decompositions of products of NS characters in the preceding
section can (by spectral flow) be used to write down the decomposition of
products of R characters. We will distinguish $c=3$-representations
corresponding to Ramond ground states by
$^{\text{g}}\Rdgch{\frac{1}{8}}{q}$, for the others we will use the
notation $\Rdgch{h}{q}$ where the charges of the lowest lying states
are $q\pm \tfrac{1}{2}$. 

\noindent The decompositions which are relevant for B-type
boundary conditions are
\begin{align}
\Rmmch{1}{1}{0}\Rmmch{4}{4}{0}\ =\ & 
^{\text{g}}\Rdgch{\frac{1}{8}}{0}+ 
\sum_{n=0}^{\infty} \Big( \Rdgch{\frac{1}{8}+3+3n}{\frac{1}{2}} +
\Rdgch{\frac{1}{8}+1+3n}{\frac{1}{2}} \nonumber \\
& \qquad \qquad \qquad \qquad 
+\Rdgch{\frac{1}{8}+2+3n}{-\frac{1}{2}} 
+ \Rdgch{\frac{1}{8}+3+3n}{-\frac{1}{2}}  \Big) \\
\Rmmch{1}{2}{0}\Rmmch{4}{2}{0}\ =\ &
^{\text{g}}\Rdgch{\frac{1}{8}}{0}+ \sum_{n=0}^{\infty}
\Big(\Rdgch{\frac{1}{8}+3+3n}{\frac{1}{2}} +
\Rdgch{\frac{1}{8}+2+3n}{\frac{1}{2}} \nonumber \\
& \qquad \qquad \qquad \qquad +\Rdgch{\frac{1}{8}+1+3n}{-\frac{1}{2}} 
+ \Rdgch{\frac{1}{8}+3+3n}{-\frac{1}{2}} \Big) \\
\Rmmch{1}{2}{0}\Rmmch{4}{3}{1}\ =\ &
\sum_{n=0}^{\infty} \Big(\Rdgch{\frac{5}{8}+3n}{\frac{1}{2}}
+\Rdgch{\frac{5}{8}+1+3n}{\frac{1}{2}} \nonumber \\
&\qquad +\Rdgch{\frac{5}{8}+1+3n}{-\frac{1}{2}} 
+ \Rdgch{\frac{5}{8}+2+3n}{-\frac{1}{2}}  \Big) \\
\Rmmch{1}{1}{1}\Rmmch{4}{2}{2}\ =\ &
\sum_{n=0}^{\infty} \Big(\Rdgch{\frac{1}{8}+1+3n}{-\frac{1}{2}}
+\Rdgch{\frac{1}{8}+2+3n}{-\frac{1}{2}} \nonumber \\
&\qquad +\Rdgch{\frac{1}{8}+1+3n}{\frac{1}{2}} 
+ \Rdgch{\frac{1}{8}+2+3n}{\frac{1}{2}} \Big) \\
\Rmmch{1}{1}{1}\Rmmch{4}{1}{1}\ =\ &
\sum_{n=0}^{\infty} \Big(\Rdgch{\frac{5}{8}+3n}{-\frac{1}{2}}
+\Rdgch{\frac{5}{8}+2+3n}{-\frac{1}{2}} \nonumber \\
&\qquad +\Rdgch{\frac{5}{8}+3n}{\frac{1}{2}} 
+ \Rdgch{\frac{5}{8}+2+3n}{\frac{1}{2}} \Big) \\
\Rmmch{1}{1}{0}\Rmmch{4}{1}{3}\ =\ &
\sum_{n=0}^{\infty} \Big(\Rdgch{\frac{5}{8}+3n}{-\frac{1}{2}}
+\Rdgch{\frac{5}{8}+1+3n}{-\frac{1}{2}} \nonumber \\
&\qquad + \Rdgch{\frac{5}{8}+1+3n}{\frac{1}{2}} 
+ \Rdgch{\frac{5}{8}+2+3n}{\frac{1}{2}} \Big) \ .
\end{align}

\subsection{Relations between minimal models with level 1 and 4}
The $N=2$ minimal model with central charge $c=1$ ($k=1$) has 3 sectors:
\begin{equation}
(\tfrac{1}{2},\tfrac{1}{2}) , (\tfrac{1}{2},\tfrac{3}{2}) ,
(\tfrac{3}{2},\tfrac{1}{2})  \ .
\end{equation}
Products of two of these characters can be decomposed in terms of the
diagonal $N=2$ algebra with central charge $c=2$:
\begin{align}
\mmch{1}{1}{1} \mmch{1}{1}{1} \ =\ & \mmch{4}{1}{1} + \mmch{4}{5}{5} \\
\mmch{1}{3}{1} \mmch{1}{1}{1} \ =\ & \mmch{4}{5}{1} \\
\mmch{1}{3}{1} \mmch{1}{3}{1} \ =\ & \mmch{4}{9}{1} + \mmch{4}{3}{7} \\
\mmch{1}{3}{1} \mmch{1}{1}{3} \ =\ & \mmch{4}{3}{3} \ .
\end{align}
From this it is easy to see that the D-type minimal model at level $4$
is equivalent to the product of two (A-type) minimal models at level
$1$. 

\noindent Let us note further relations between characters:
\begin{align}
\mmch[z^{3},q^{3}]{1}{1}{1} \ =\ &
\mmch{1}{1}{1}\big( \mmch{4}{1}{1} + \mmch{4}{5}{5} \big)  \nonumber\\
& \qquad - 3 \sum_{m=1}^{\infty}\sum_{\substack{n=1\\
\makebox[0cm][c]{$\scriptstyle\!\!\!\!\!m+n \not\equiv 0
\,\text{mod}\, 3$}}}^{m} \ \dgch{m^{2}+n^{2}-mn}{0} \\
\mmch[z,q^{\frac{1}{3}}]{1}{3}{1}\ =\
& \sum_{n=0}^{\infty} \Big( \dgch{\frac{1+2n}{6}}{\frac{1}{3}} +
\dgch{\frac{1+2n}{3}}{-\frac{2}{3}} \Big) \nonumber\\
=\ & 
\mmch{1}{1}{1}\mmch{4}{5}{1} + \mmch{1}{1}{3}\mmch{4}{1}{5} \nonumber\\
& \qquad +\mmch{1}{3}{1}\mmch{4}{3}{3} \\
\mmch[z,q^{\frac{1}{3}}]{1}{1}{1}\
=\ &\sum_{m=1}^{\infty}\sum_{\substack{n=1\\ 
\makebox[0cm][c]{$\scriptstyle\!\!\!\!\!m+n \not\equiv 0
\,\text{mod}\, 3$}}}^{m} \ \dgch{\frac{m^{2}+n^{2}-mn}{9}}{0} \ .
\end{align}

\subsection{Some character identities}

We use the usual conventions for the theta functions:
\begin{align}
\vartheta_{1} (z,q) & = -i \, q^{1/12}\, \eta (q) \,
(z^{1/2}-z^{-1/2})\, \prod_{n=1}^{\infty} (1-z\, q^{n})
(1-z^{-1}\, q^{n})  \nonumber \\
\vartheta_{2}(z,q) & = q^{1/12}\, \eta(q) \, (z^{1/2}+z^{-1/2}) \,
\prod_{n=1}^{\infty} (1+z\, q^{n}) (1+z^{-1}\, q^{n})
\nonumber \\
\vartheta_{3}(z,q) & = q^{-1/24}\, \eta(q) \,  \prod_{n=1}^{\infty}
(1+z\, q^{n-1/2}) (1+z^{-1}\, q^{n-1/2}) \nonumber \\
\vartheta_{4}(z,q) & = q^{-1/24}\, \eta(q) \,  
\prod_{n=1}^{\infty}
(1-z\, q^{n-1/2}) (1-z^{-1}\, q^{n-1/2})\nonumber \\
\eta(q) & = q^{1/24} \prod_{n=1}^{\infty} (1 - q^n) \ . \nonumber 
\end{align}
Finally, we collect some identities relating theta functions to 
characters of the $k=1$ and $k=4$ $N=2$ minimal models:
\begin{align}
\frac{1}{2}\sum_{p\in
\Lambda_{R}}\tilde{q}^{\frac{1}{2}p^{2}}\frac{\vartheta_{3}
(\tilde{q})}{\eta^{3} (\tilde{q})} 
= & \frac{1}{2} \Big( \mmch[\tilde{q}]{1}{1}{1} 
\big( \mmch[\tilde{q}]{4}{1}{1} +\mmch[\tilde{q}]{4}{5}{5} \big) \nonumber\\
& +\mmch[\tilde{q}]{1}{3}{1} 
\big( \mmch[\tilde{q}]{4}{9}{1} +\mmch[\tilde{q}]{4}{3}{7}\big) \nonumber\\
& +\mmch[\tilde{q}]{1}{1}{3} 
\big( \mmch[\tilde{q}]{4}{1}{9} +\mmch[\tilde{q}]{4}{7}{3} \big)  \Big)\\
\frac{1}{2} \sum_{p\in \Lambda_{R}}
\tilde{q}^{\frac{1}{2} (p+u_{1})^{2}} 
\frac{\vartheta_{3} (\tilde{q})}{\eta^{3}
(\tilde{q})} 
= & \sum_{m=1}^{\infty}\sum_{\substack{n=1\\
\makebox[0cm][c]{$\scriptstyle\mspace{-50mu}m\ \text{or}\ n\
\text{odd}$}}}^{m} \dgch[\tilde{q}]{\frac{m^{2}+n^{2}-mn}{4}}{0}
\end{align}
\begin{align}
\frac{\vartheta_{3} (\tilde{q})}{\vartheta_{2} (\tilde{q})}
= & \frac{1}{2} \Big(\tmmch{1}{1}{1} 
\big( \tmmch{4}{1}{1} -\tmmch{4}{5}{5} \big) \nonumber\\
&+\tmmch{1}{3}{1} 
\big(\tmmch{4}{9}{1} +\tmmch{4}{3}{7} \big) \nonumber\\
&+\tmmch{1}{1}{3} 
\big(\tmmch{4}{1}{9} +\tmmch{4}{7}{3} \big)  \Big)\\
-i\frac{\vartheta_{3}
(\tilde{q}^{-\frac{1}{3}},\tilde{q})}{\vartheta_{1}
(\tilde{q}^{-\frac{1}{3}},\tilde{q})} 
= & \frac{1}{2} \Big( \mmch[\tilde{q}]{1}{1}{1}
\big(\mmch[\tilde{q}]{4}{9}{1} + \mmch[\tilde{q}]{4}{7}{3} +
\mmch[\tilde{q}]{4}{3}{7} + \mmch[\tilde{q}]{4}{1}{9}\big) \nonumber\\
&+  \mmch[\tilde{q}]{1}{3}{1}
\big(\mmch[\tilde{q}]{4}{1}{1} + \mmch[\tilde{q}]{4}{5}{5} +
\mmch[\tilde{q}]{4}{7}{3} + \mmch[\tilde{q}]{4}{1}{9}\big) \nonumber\\
&+  \mmch[\tilde{q}]{1}{1}{3}
\big(\mmch[\tilde{q}]{4}{1}{1} + \mmch[\tilde{q}]{4}{5}{5} +
\mmch[\tilde{q}]{4}{3}{7} + \mmch[\tilde{q}]{4}{9}{1}\big)  \Big)\\
\frac{\vartheta_{3}
(\tilde{q}^{-\frac{1}{3}},\tilde{q})}{\vartheta_{2}
(\tilde{q}^{-\frac{1}{3}},\tilde{q})} 
= & \frac{1}{2} \Big( \tmmch{1}{1}{1}
\big(\tmmch{4}{9}{1} + \tmmch{4}{7}{3} +
\tmmch{4}{3}{7} + \tmmch{4}{1}{9}\big) \nonumber\\
& +  \tmmch{1}{3}{1}
\big(\tmmch{4}{1}{1} - \tmmch{4}{5}{5} -
\tmmch{4}{7}{3} - \tmmch{4}{1}{9}\big) \nonumber\\
& +  \tmmch{1}{1}{3}
\big(\tmmch{4}{1}{1} - \tmmch{4}{5}{5} -
\tmmch{4}{3}{7} - \tmmch{4}{9}{1}\big)  \Big)\\
\tilde{z}^{-\frac{1}{3}} 
\frac{\vartheta_{3}(\tilde{z}\tilde{q}^{-1/3},\tilde{q})}{\vartheta_{2}
(\tilde{q}^{-1/3},\tilde{q})} & -\tilde{z}^{\frac{1}{3}}
\frac{\vartheta_{3}(\tilde{z}^{-1}\tilde{q}^{-1/3},\tilde{q})}{\vartheta_{2}
(\tilde{q}^{-1/3},\tilde{q})}\nonumber\\
=& \tmmch[\tilde{z},\tilde{q}]{1}{1}{3} 
\Big( \tmmch[\tilde{z},\tilde{q}]{4}{1}{1} +
\tmmch[\tilde{z},\tilde{q}]{4}{9}{1} 
- \tmmch[\tilde{z},\tilde{q}]{4}{5}{5} 
+ \tmmch[\tilde{z},\tilde{q}]{4}{3}{7} \Big) \nonumber\\
& - \tmmch[\tilde{z},\tilde{q}]{1}{3}{1} 
\Big( \tmmch[\tilde{z},\tilde{q}]{4}{1}{1} 
+ \tmmch[\tilde{z},\tilde{q}]{4}{1}{9}
- \tmmch[\tilde{z},\tilde{q}]{4}{5}{5} 
+ \tmmch[\tilde{z},\tilde{q}]{4}{7}{3} \Big)\nonumber\\
& - \tmmch[\tilde{z},\tilde{q}]{1}{1}{1} \Big(
\tmmch[\tilde{z},\tilde{q}]{4}{1}{9} 
+\tmmch[\tilde{z},\tilde{q}]{4}{7}{3} 
-\tmmch[\tilde{z},\tilde{q}]{4}{3}{7} 
-\tmmch[\tilde{z},\tilde{q}]{4}{9}{1} \Big) \ .
\end{align}

\section{Matrix factorisations of the permutation orbifold}

In order to understand the relation between the D-branes of the
$A_1\times A_4$ model and the orbifold of the 
$A_1\times A_1\times A_1'$ theory, we now study some simple matrix
factorisations of the latter theory. 
The simplest factorisation is of permutation type in the factors $v,w$
and tensor in the rest, 
\begin{equation}
Q_{u(vw)z;\xi} = \left(\begin{matrix}0 & u\\ u^{2}& 0\end{matrix}\right) 
\hat{\otimes} \ 
Q_{(vw);\xi} \ \hat{\otimes} \ Q_{z} \ .
\end{equation}
Here, $Q_{z}$ is the simplest factorisation of the $z$-part, while 
$Q_{(vw);\xi}$ is the permutation factorisation, 
\begin{equation}
Q_{z} = \left(\begin{matrix}0&z\\ z&0  \end{matrix} \right) \ , \qquad 
Q_{(vw);\xi} = 
\left( \begin{matrix}0 & (v-\xi w)\\ \pi_{\xi} (v,w) & 0
\end{matrix}\right) \ ,
\end{equation}
where $\pi_\xi$ is defined as in (\ref{defofpi}). The symbol
$\hat{\otimes}$ denotes the tensoring of matrix factorisations
\cite{Ashok:2004zb} that was explained in some detail in
\cite{Hori:2004ja}.   

\noindent The behaviour of this factorisation under the 
$\mathbb{Z}_{2}$-orbifold action depends on the value of $\xi$. If
$\xi=-1$, then $Q$ transforms under the $\mathbb{Z}_{2}$-action 
(that exchanges $v$ and $w$ and maps $z\mapsto -z$) 
to an equivalent factorisation $Q'$, 
\begin{equation}
Q_{u (vw)z;\xi=-1} = \omega \ Q'_{u (vw)z;\xi=-1} \ \omega^{-1}\ .
\end{equation}
There are two choices for $\omega$ which we denote by $\omega_\pm$, 
\begin{equation}
\omega_{\pm} = \pm \mathbf{1}_{2}\otimes  
\mathbf{1}_{2}
\otimes \left( \begin{matrix}1 & 0 \\ 0& -1 \end{matrix}\right) \ .
\end{equation}
In order to determine the self-spectrum of this factorisation we note
that before taking the orbifold, we have 4 bosons and 4 fermions.
In the orbifold theory we are only allowed to keep the 
equivariant morphisms, so we have to determine how the morphisms
coming from the different factors behave under the
$\mathbb{Z}_{2}$-action.  From the $u$-part there is one boson and one
fermion, both of which are even under the $\mathbb{Z}_{2}$-action. The
$(vw)$-part has 2 bosons of which one is even and the other is odd; it
does not have any fermions. The $z$-part has one even boson and one
odd fermion. Tensoring those together and restricting to the overall
even morphisms we find in total 2 bosons and 2 fermions. This is the
same spectrum that we have found for the $T(0,0)$ tensor brane in the 
$A_{1}\times A_{4}$-model. The two choices $\omega_{\pm}$ 
correspond to brane and anti-brane.

For $\xi=e^{\pm \frac{i\pi}{6}}$, the image under the orbifold action 
$Q'_{u (vw)z;\xi}$ is equivalent to $Q_{u (vw)z;\bar\xi}$; in order to
obtain an equivariant factorisation we therefore have to take the 
superposition of the two factorisations. This leads then to the
$T(0,2)$ tensor brane in the $A_{1}\times A_{4}$-model.   

Finally, in order to obtain the $T(0,1)$-brane we consider the 
tensor product factorisation in the $A_{1}\times A_1 \times A_1'$
model  
\begin{equation}
Q_{\text{t}} = 
\left(\begin{matrix} 0&u\\ u^{2} & 0\end{matrix}\right) \hat{\otimes} 
\left( \begin{matrix}0&0&v&w\\ 0&0&-w^{2} & v^{2}\\
v^{2}&-w&0&0\\
w^{2}&v&0&0\end{matrix}\right) \hat{\otimes} \ Q_{z}  \quad
\hbox{with} \quad
\omega_{\pm}  = \pm \mathbf{1}_{2} \otimes 
\left( \begin{matrix}1&0&0&0\\ 0&-1&0&0\\ 0&0&0&1\\ 0&0&1&0 
\end{matrix}\right)
\otimes \left( \begin{matrix}1&0\\ 0& -1\end{matrix}\right) \ .
\end{equation}
From the $v,w$-part we get two even bosons and two odd
fermions. Tensoring this with the other morphisms we find in total
four bosons and four fermions. This is indeed what we expect for the
$T(0,1)$ tensor brane in the $A_{1}\times A_{4}$-model (see e.g.\ 
\cite{Brunner:2005fv}). The two choices of $\omega $ correspond to
brane and anti-brane.

\subsection[The matrix factorisations for $\hat{P}$ and $P$]{The 
matrix factorisations for $\boldsymbol{\hat{P}}$ and $\boldsymbol{P}$}

This accounts for the tensor product branes of the 
$A_{1}\times A_{4}$-model. The $\hat{P}$-branes are now 
related to factorisations of the 
$A_{1}\times A_1 \times A_1'$ theory that involve a permutation of the 
$u$-factor. The simplest such factorisation that is 
equivariant with respect to the $\mathbb{Z}_{2}$-action involves a
superposition of a $(uv)$-permutation and a $(uw)$-permutation  with the 
same choice for $\xi $,
\begin{equation}
Q_{\hat{P}} = \left(\begin{matrix}
0&0&u-\xi v & w\\
0&0&-w^{2}   & \pi_{\xi} (u,v)\\
\pi_{\xi} (u,v) & -w&0&0\\
w^{2} & u-\xi v &0&0
\end{matrix}\right) \hat{\otimes} Q_{z} 
\oplus 
\Bigl( v \leftrightarrow w \Bigr) \ .
\end{equation}
The equivariance map $\omega$ can then be chosen to 
permute the two factorisations in the superposition. Before
orbifolding we have four bosons and four fermions coming from each of
the two self-spectra, and four bosons and four fermions from the
relative overlap, giving in total  12 bosons and 12 fermions. Half of
those are even under the $\mathbb{Z}_{2}$-action, so we find 6 bosons
and 6 fermions in the  orbifold. This agrees with the results of
(\ref{nn1}). 
\smallskip

This begs the question of what the $P(j)$-branes of the 
$A_{1}\times A_{4}$ model correspond to in the 
$A_1\times A_1\times A_1'$ theory. We propose that they correspond to
the factorisation  
\begin{equation}
Q_{P} = \left(\begin{matrix} 0&E_{P} \\ J_{P} &0  \end{matrix} \right)
\hat{\otimes} Q_{z}
\end{equation}
with
\begin{align}
E_{P} = &\left(\begin{matrix}\pi_{\xi} (u',v) & u'-\xi w\\
\pi_{\xi} (u',w)& - (u'-\xi v)  \end{matrix} \right) &
J_{P} = &\left(\begin{matrix}u'-\xi v & u'-\xi w\\
\pi_{\xi} (u',w) & -\pi_{\xi} (u',v)  \end{matrix} \right)\ .
\end{align}
Here, we have rescaled $u$ to $u'=u/\sqrt[3]{2}$, and the three
different values of $j$ correspond to the three different values of
$\xi$. The representation $\omega$ of the orbifold group is 
\begin{equation}
\omega = \left(\begin{matrix}\omega_{0} &0\\
0&\omega_{1}\end{matrix}\right) \otimes
\left(\begin{matrix}1&0\\ 0& -1 \end{matrix} \right) \ , \qquad
\hbox{with} \qquad 
\omega_{0} = \left(\begin{matrix}0&1\\ 1&0 \end{matrix} \right) \ ,
\qquad 
\omega_{1} = \left(\begin{matrix}1&0\\ 0& -1 \end{matrix} \right) \ .
\end{equation}
In order to support this claim we shall now determine the
self-spectrum of this factorisation. 

\subsubsection[The spectrum of the $P$-factorisation]{The 
spectrum of the $\boldsymbol{P}$-factorisation}

First we note that $Q_{P}$ has the structure of a tensor product of a  
permutation factorisation of the $(u'v)$- and one of the
$(u'w)$-part. The spectrum, however, is not just the product of the two
spectra, because the variables involved in the two factorisations are
not independent.  To begin with, let us ignore the $z$-part of
$Q_{P}$. A closed boson
\begin{equation}
\phi = \left(\begin{matrix}\phi_{0}&0\\ 0&\phi_{1}  \end{matrix} \right)
\end{equation}
satisfies $E_{P}\phi_{1}=\phi_{0}E_{P}$ and $J_{P}\phi_{0}=\phi_{1}J_{P}$.
By adding exact morphisms, it can always be brought to the form
\begin{equation}
\phi_{0} = \phi_{1} =  \left(\begin{matrix}s & 0\\ 0& s  
\end{matrix} \right) 
\end{equation}
where $s$ is a polynomial in $u',v,w$. The remaining freedom of adding
exact morphisms means that $s$ is defined up to elements in the ideal
generated by $(u'-\xi v)$, $(u'-\xi w)$, $\pi_{\xi} (u',v)$ and
$\pi_{\xi}(u',w)$. We can easily see that two bosons remain, both of
which are even under the action of the orbifold group.

\noindent A closed fermion
\begin{equation}
\psi = \left(\begin{matrix}0&\psi_{1} \\\psi_{0} & 0  \end{matrix} \right)
\end{equation}
satisfies $E_{P}\psi_{0}+\psi_{1}J_{P}=0$ and
$J_{P}\psi_{1}+\psi_{0}E_{P}=0$. By adding exact morphisms, it can
always be brought to the form
\begin{align}
\psi_{0} = & \left(\begin{matrix}t & -t \\ 
t (u'+\xi v+\xi w) & t (u'+\xi v+\xi w)\end{matrix} \right)\, , &\, 
\psi_{1} = & \left(\begin{matrix}-t (u'+\xi v+\xi w) & -t\\
t (u'+\xi v+\xi w) & -t  \end{matrix} \right) \ .
\end{align}
The remaining freedom of adding exact morphisms means that the
polynomial $t$ is defined up to elements in the ideal generated by 
$(u'-\xi v)$, $(u'-\xi w)$ and $(v^{2}+vw+w^{2})$. There are two
fermions that remain which can be represented by $t=1$ and
$t=v+w$. Under the orbifold action, these two fermions transform as  
\begin{equation}
\psi (u,v,w) \to \omega \psi (u,w,v) \omega^{-1} = -\psi (u,v,w) \ ,
\end{equation} 
so they are odd under the $\mathbb{Z}_{2}$-action.

Tensoring the spectrum together with the even boson and the odd
fermion coming from the $z$-part, we find four bosons and no fermions
in the orbifolded theory. This then agrees with the open string
spectrum on the generalised permutation factorisation of the 
$A_{1}\times A_{4}$-model.

\providecommand{\href}[2]{#2}\begingroup\raggedright
\endgroup


\begin{thebibliography}{10}

\bibitem{Gepner:1987qi}
D.~Gepner, {\it Space-time supersymmetry in compactified string theory and
superconformal models}, 
{\em Nucl.\ Phys. } {\bf B296} (1988) 757. 
%%CITATION = NUPHA,B296,757;%%

\bibitem{Vafa1}
C.~Vafa,
{\it C theorem and the topology of 2-d QFTs}, 
{\em Phys.\ Lett. } {\bf B212} (1988) 27.
%%CITATION = PHLTA,B212,28;%%

\bibitem{Mar1}
E.J.~Martinec,
{\it Algebraic geometry and effective Lagrangians},
{\em Phys.\ Lett. } {\bf B217} (1989) 431.
%%CITATION = PHLTA,B217,431;%%

\bibitem{VW}
C.~Vafa and N.P.~Warner,
{\it Catastrophes and the classification of conformal theories}, 
{\em Phys.\ Lett. } {\bf B218} (1989) 51.
%%CITATION = PHLTA,B218,51;%%

\bibitem{W}
E.~Witten,
{\it On the Landau-Ginzburg description of $N=2$ minimal models}, 
{\em Int.\ Journ.\ Mod.\ Phys. } {\bf A9} (1994) 4783
[\href{http://arXiv.org/abs/hep-th/9304026}{{\tt hep-th/9304026}}].
%%CITATION = HEP-TH 9304026;%%

\bibitem{Witten:1993yc}
E.~Witten,
{\it Phases of N = 2 theories in two dimensions},
{\em Nucl.\ Phys. } {\bf B403} (1993) 159
[\href{http://arXiv.org/abs/hep-th/9301042}{{\tt hep-th/9301042}}].
%%CITATION = HEP-TH 9301042;%%

\bibitem{Recknagel:1998sb}
A.~Recknagel and V.~Schomerus, {\it D-branes in {Gepner} models},
{\em Nucl.\ Phys. } {\bf B531} (1998) 185
[\href{http://arXiv.org/abs/hep-th/9712186}{{\tt hep-th/9712186}}].
%%CITATION = HEP-TH 9712186;%%

\bibitem{Brunner:1999jq}
I.~Brunner, M.R.~Douglas, A.E.~Lawrence and C.~R\"omelsberger,
{\it D-branes on the quintic},
{\em JHEP} {\bf 08} (2000) 015
[\href{http://arXiv.org/abs/hep-th/9906200}{{\tt hep-th/9906200}}].
%%CITATION = HEP-TH 9906200;%%

\bibitem{Recknagel:2002qq}
A.~Recknagel, {\it Permutation branes},  
{\em JHEP} {\bf 04} (2003) 041
[\href{http://arXiv.org/abs/hep-th/0208119}{{\tt hep-th/0208119}}].
%%CITATION = HEP-TH 0208119;%%

\bibitem{Brunner:2005fv}
I.~Brunner and M.R. Gaberdiel, {\it Matrix factorisations and permutation
branes},  
{\em JHEP} {\bf 07} (2005) 012
[\href{http://arXiv.org/abs/hep-th/0503207}{{\tt hep-th/0503207}}].
%%CITATION = HEP-TH 0503207;%%

\bibitem{Caviezel:2005th}
C.~Caviezel, S.~Fredenhagen and M.R.~Gaberdiel,
{\it The RR charges of A-type Gepner models}, 
{\em JHEP} {\bf 01} (2006) 111
[\href{http://arXiv.org/abs/hep-th/051107}{{\tt hep-th/051107}}].
%%CITATION = HEP-TH 0051107;%%

\bibitem{Kapustin:2002bi}
A.~Kapustin and Y.~Li, {\it D-branes in Landau-Ginzburg models and
algebraic geometry},  
{\em JHEP} {\bf 12} (2003) 005
[\href{http://arXiv.org/abs/hep-th/0210296}{{\tt hep-th/0210296}}].
%%CITATION = HEP-TH 0210296;%%

\bibitem{Brunner:2003dc}
I.~Brunner, M.~Herbst, W.~Lerche and B.~Scheuner, {\it Landau-Ginzburg
realization of open string TFT},
\href{http://arXiv.org/abs/hep-th/0305133}{{\tt hep-th/0305133}}.
%%CITATION = HEP-TH 0305133;%%

\bibitem{KL2}
A.~Kapustin and Y.~Li,
{\it Topological correlators in Landau-Ginzburg models with
boundaries}, 
{\em Adv.\ Theor.\ Math.\ Phys. }  {\bf 7} (2004) 727
[\href{http://arXiv.org/abs/hep-th/0305136}{{\tt hep-th/0305136}}].
%%CITATION = HEP-TH 0305136;%%

\bibitem{Lazaroiu:2003zi}
C.I.~Lazaroiu,
{\it On the boundary coupling of topological Landau-Ginzburg models},
{\em JHEP} {\bf 05} (2005) 037
[\href{http://arXiv.org/abs/hep-th/0312286}{{\tt hep-th/0312286}}].
%%CITATION = HEP-TH 0312286;%%

\bibitem{Herbst:2004ax}
M.~Herbst and C.I.~Lazaroiu,
{\it Localization and traces in open-closed topological 
Landau-Ginzburg  models}, 
{\em JHEP} {\bf 05} (2005) 044
[\href{http://arXiv.org/abs/hep-th/0404184}{{\tt hep-th/0404184}}].
%%CITATION = HEP-TH 0404184;%%

\bibitem{Ashok:2004zb}
S.K. Ashok, E.~Dell'Aquila and D.-E.~Diaconescu, {\it Fractional
branes in Landau-Ginzburg orbifolds},  
{\em Adv.\ Theor.\ Math.\ Phys. } {\bf 8} (2004) 461
[\href{http://arXiv.org/abs/hep-th/0401135}{{\tt hep-th/0401135}}]. 
%%CITATION = HEP-TH 0401135;%%

\bibitem{Ashok:2004xq}
S.K.~Ashok, E.~Dell'Aquila, D.-E.~Diaconescu and B.~Florea,
{\it Obstructed D-branes in Landau-Ginzburg orbifolds},
{\em Adv.\ Theor.\ Math.\ Phys. }  {\bf 8} (2004) 427
[\href{http://arXiv.org/abs/hep-th/0404167}{{\tt hep-th/0404167}}]. 
%%CITATION = HEP-TH 0404167;%%

\bibitem{Hori:2004ja}
K.~Hori and J.~Walcher, {\it F-term equations near Gepner points},
{\em JHEP} {\bf 01} (2005) 008
[\href{http://arXiv.org/abs/hep-th/0404196}{{\tt  hep-th/0404196}}].
%%CITATION = HEP-TH 0404196;%%

\bibitem{Enger:2005jk}
H.~Enger, A.~Recknagel and D.~Roggenkamp,
{\it Permutation branes and linear matrix factorisations},
{\em JHEP} {\bf 01} (2006) 087
[\href{http://arXiv.org/abs/hep-th/0508053}{{\tt hep-th/0508053}}].
%%CITATION = HEP-TH 0508053;%%

\bibitem{Klemm:2003vn}
H.~Klemm,
{\it Embedding diagrams of the N = 2 superconformal algebra under
spectral flow}, 
{\em Int.\ J.\ Mod.\ Phys. }  {\bf A19} (2004) 5263
[\href{http://arXiv.org/abs/hep-th/0306073}{{\tt hep-th/0306073}}].
%%CITATION = HEP-TH 0306073;%%

\bibitem{Gaberdiel:2004nv}
M.R.~Gaberdiel and H.~Klemm,
{\it N = 2 superconformal boundary states for free bosons and
fermions}, 
{\em Nucl.\ Phys. }  {\bf B693} (2004) 281
[\href{http://arXiv.org/abs/hep-th/0404062}{{\tt hep-th/0404062}}]. 
%%CITATION = HEP-TH 0404062;%%

\bibitem{Lerche:1989cs}
W.~Lerche, D.~L{\"u}st and N.P.~Warner,
{\it Duality symmetries in N=2 Landau-Ginzburg models},
{\em Phys.\ Lett. } {\bf B231} (1989) 417.
%%CITATION = PHLTA,B231,417;%%

\bibitem{Fuchs:2000gv}
J.~Fuchs, C.~Schweigert and J.~Walcher,
{\it Projections in string theory and boundary states for Gepner
models}, 
{\em Nucl.\ Phys. }  {\bf B588} (2000) 110
[\href{http://arXiv.org/abs/hep-th/0003298}{{\tt hep-th/0003298}}].
%%CITATION = HEP-TH 0003298;%%

\bibitem{Hori:2004zd}
K.~Hori and J.~Walcher, {\it D-branes from matrix factorizations},
{\em Comptes Rendus Physique} {\bf 5} (2004) 1061
[\href{http://arXiv.org/abs/hep-th/0409204}{{\tt hep-th/0409204}}].
%%CITATION = HEP-TH 0409204;%%

\bibitem{Kapustin:2003rc}
A.~Kapustin and Y.~Li, {\it D-branes in topological minimal models: The
Landau-Ginzburg approach},  
{\em JHEP} {\bf 07} (2004) 045
[\href{http://arXiv.org/abs/hep-th/0306001}{{\tt hep-th/0306001}}].
%%CITATION = HEP-TH 0306001;%%

\bibitem{Gaberdiel:2002jr}
M.R.~Gaberdiel and S.~Sch\"afer-Nameki,
{\it D-branes in an asymmetric orbifold},
{\em  Nucl.\ Phys. }  {\bf B654} (2003) 177
[\href{http://arXiv.org/abs/hep-th/0210137}{{\tt hep-th/0210137}}].
%%CITATION = HEP-TH 0210137;%%

\bibitem{Sarkissian:2003yw}
G.~Sarkissian and M.~Zamaklar,
{\it Symmetry breaking, permutation D-branes on group manifolds:
Boundary states and geometric description},
{\em Nucl.\ Phys. } {\bf B696} (2004) 66
[\href{http://arXiv.org/abs/hep-th/0312215}{{\tt hep-th/0312215}}]. 
%%CITATION = HEP-TH 0312215;%%


\bibitem{Dell'Aquila:2005jg}
E.~Dell'Aquila,
{\it D-branes in toroidal orbifolds and mirror symmetry},
{\em JHEP} {\bf 04} (2006) 035
[\href{http://arXiv.org/abs/hep-th/0512051}{{\tt hep-th/0512051}}].
%%CITATION = HEP-TH 0512051;%%

\bibitem{Brunner:2006tc}
I.~Brunner, M.R.~Gaberdiel and C.A.~Keller,
{\it Matrix factorisations and D-branes on K3},
\href{http://arXiv.org/abs/hep-th/0603196}{{\tt hep-th/0603196}}.
%%CITATION = HEP-TH 0603196;%%

\bibitem{Lewellen:1991tb}
D.C.~Lewellen,
{\it Sewing constraints for conformal field theories on surfaces with 
boundaries},
{\em Nucl.\ Phys. }  {\bf B372} (1992) 654.
%%CITATION = NUPHA,B372,654;%%

\bibitem{Recknagel:1998ih}
A.~Recknagel and V.~Schomerus,
{\it Boundary deformation theory and moduli spaces of D-branes},
{\em Nucl.\ Phys. }  {\bf B545} (1999) 233
[\href{http://arXiv.org/abs/hep-th/9811237}{{\tt hep-th/9811237}}].
%%CITATION = HEP-TH 9811237;%%

\bibitem{Fuchs:1997kt}
J.~Fuchs and C.~Schweigert,
{\it A classifying algebra for boundary conditions},
{\em Phys.\ Lett. }  {\bf B414} (1997) 251
[\href{http://arXiv.org/abs/hep-th/9708141}{{\tt hep-th/9708141}}].
%%CITATION = HEP-TH 9708141;%%

\bibitem{Gaberdiel:2001xm}
M.R.~Gaberdiel, A.~Recknagel and G.M.T.~Watts,
{\it The conformal boundary states for SU(2) at level 1},
{\em Nucl.\ Phys. }  {\bf B626} (2002) 344
[\href{http://arXiv.org/abs/hep-th/0108102}{{\tt hep-th/0108102}}].
%%CITATION = HEP-TH 0108102;%%

\bibitem{Brunner:2005pq}
I.~Brunner and M.R.~Gaberdiel,
{\it The matrix factorisations of the D-model},
{\em J.\ Phys. } {\bf A38} (2005) 7901
[\href{http://arXiv.org/abs/hep-th/0506208}{{\tt hep-th/0506208}}].
%%CITATION = HEP-TH 0506208;%%


\bibitem{Fredenhagen:2005an}
S.~Fredenhagen and T.~Quella,
{\it Generalised permutation branes},
{\em JHEP} {\bf 11} (2005) 004
[\href{http://arXiv.org/abs/hep-th/0509153}{{\tt hep-th/0509153}}].
%%CITATION = HEP-TH 0509153;%%

\bibitem{Sarkissian:2006xp}
G.~Sarkissian,
{\it Generalised permutation branes on a product of cosets 
G(k(1))/H $\times$ G(k(2))/H},
{\em Nucl.\ Phys. }  {\bf B747} (2006) 423
[\href{http://arXiv.org/abs/hep-th/0601061}{{\tt hep-th/0601061}}].
%%CITATION = HEP-TH 0601061;%%


\end{thebibliography}
\end{document}